\documentclass[11pt]{article}

\usepackage[preprint]{acl}

\usepackage{times}
\usepackage{latexsym}
\usepackage{amsfonts}
\usepackage{amsmath}
\usepackage{amssymb}
\usepackage{calligra}
\usepackage{mathrsfs}
\usepackage{algorithm}
\usepackage{algorithmicx}
\usepackage{algpseudocode}
\usepackage{booktabs}
\usepackage{multirow}
\usepackage[table]{xcolor}
\usepackage{enumitem}
\usepackage{makecell}

\usepackage[T1]{fontenc}

\usepackage[utf8]{inputenc}

\usepackage{microtype}

\usepackage{inconsolata}

\usepackage{graphicx}

\title{ FineLAP: Taming Heterogeneous Supervision for Fine-grained Language-Audio Pretraining }

\author{
 \textbf{Xiquan Li\textsuperscript{1,2}},
 \textbf{Xuenan Xu\textsuperscript{1}},
 \textbf{Ziyang Ma\textsuperscript{1}},
 \textbf{Wenxi Chen\textsuperscript{1,3}},
 \textbf{Haolin He\textsuperscript{4}}, \\
 \textbf{Qiuqiang Kong\textsuperscript{4}},
 \textbf{Xie Chen\textsuperscript{1,3}},
\\
 \textsuperscript{1}X-LANCE Lab, Shanghai Jiao Tong University, China, \\
 \textsuperscript{2}SJTU Paris Elite Institute of Technology, Shanghai Jiao Tong University, China,  \\
 \textsuperscript{3}Shanghai Innovation Institute, China, \textsuperscript{4}The Chinese University of Hong Kong, China
\\
 \small{
   \href{mtxiaoxi55@sjtu.edu.cn}{mtxiaoxi55@sjtu.edu.cn}; \href{chenxie95@sjtu.edu.cn}{chenxie95@sjtu.edu.cn}
 }
}

\begin{document}
\maketitle

\begin{abstract}
Contrastively pretrained audio–language models (e.g., CLAP) excel at clip-level understanding but struggle with frame-level tasks.
Existing extensions fail to exploit the varying granularity of real-world audio–text data, where massive clip-level textual descriptions coexist with limited frame-level annotations. 
This paper proposes \textbf{Fine}-grained \textbf{L}anguage-\textbf{A}udio \textbf{P}retraining (\textbf{FineLAP}), a novel training paradigm that advances both clip- and frame-level alignment in CLAP with heterogeneous data.
FineLAP introduces a dual-stream sigmoid loss with a cluster-based sampling strategy to jointly learn from clip- and frame-level supervision. 
To capture both global semantics and local details, FineLAP uses a decoupled audio projector on top of a self-supervised encoder.
To alleviate the scarcity of temporally annotated data, we present FineLAP-100k, a large-scale synthetic SED dataset constructed through a scalable curation pipeline.
Extensive experiments demonstrate that FineLAP achieves SOTA performance across multiple audio understanding tasks, including retrieval, classification, sound event detection, and text-to-audio grounding. 
Ablation studies further show that coarse- and fine-grained alignment are mutually beneficial, providing insights for building better audio-language models (ALMs).
\footnote{Code at: \url{https://github.com/xiquan-li/FineLAP} Dataset at: \url{https://huggingface.co/datasets/AndreasXi/FineLAP-100k}}
\end{abstract}



\section{Introduction}
Contrastively pre-trained audio–language models (e.g., CLAP) \cite{wu2023large} learn rich multimodal representations. By aligning audio and text into a shared space, CLAP facilitates a wide range of downstream tasks, including audio–text retrieval \cite{oncescu2024dissecting}, audio classification \cite{seth2024pat}, automated audio captioning \cite{li2024drcap, chen2024slam}, and text-to-audio generation \cite{liu2023audioldm, li2025meanaudio}.

\begin{figure}[t]
    \centering
    \includegraphics[width=\linewidth]{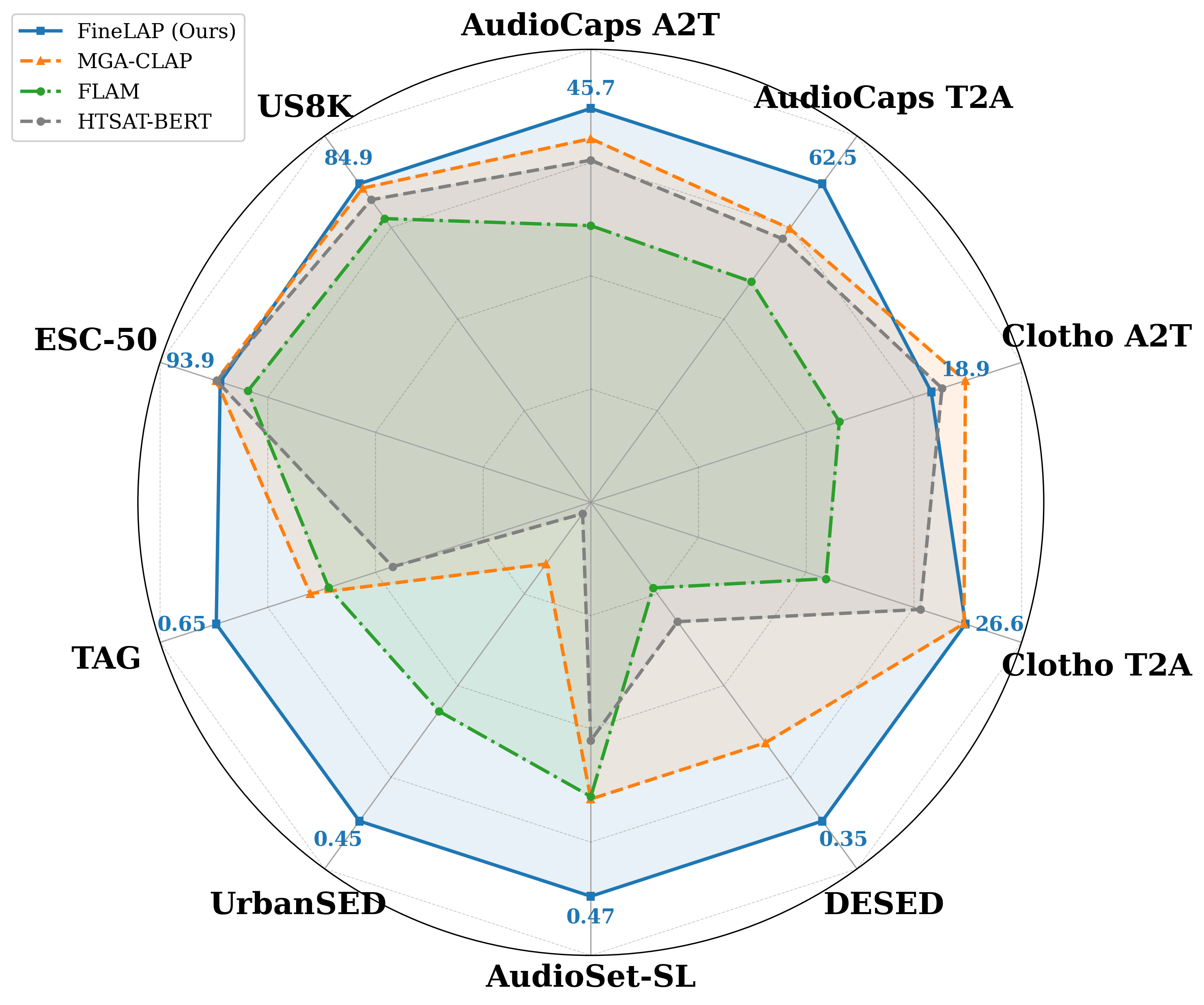}
    \caption{FineLAP’s performance across different benchmark datasets: Our model achieves state-of-the-art (SOTA) results on both clip- and frame-level tasks. }
    \label{fig:radar_performance}
    \vspace{-0.5cm}
\end{figure}


However, CLAP’s vanilla contrastive objective is defined at the clip level, where an entire audio sequence is aggregated into a single global vector and aligned to its corresponding caption. 
While such global alignment suffices for tasks like retrieval or classification, it remains suboptimal for frame-level applications. 
In such cases, establishing a fine-grained alignment between individual audio frames and textual phrases is indispensable.
This local alignment not only enhances the performance of CLAP in open-vocabulary sound event detection, but also facilitates various downstream tasks, including automated temporal annotation, reward modeling for controllable generation, and precise text-guided audio editing.
Moreover, frame-level alignment can reciprocally improve CLAP's clip-level performance through deeper cross-modal interactions. 
Nevertheless, a primary obstacle to achieving such detailed alignment lies in the prohibitive cost of obtaining temporally annotated data. 
While coarse-grained clip-level audio captions are relatively abundant, precise frame-level labels remain scarce. To overcome this bottleneck, it becomes promising to leverage heterogeneous supervision by jointly training on massive audio–captions pairs and limited frame-level annotated samples, thereby synergistically advancing both coarse- and fine-grained alignment.

In this paper, we propose \textbf{Fine}-grained \textbf{L}anguage \textbf{A}udio \textbf{P}retraining (\textbf{FineLAP}), a novel training paradigm that advances both global and fine-grained alignment in CLAP under heterogeneous supervision from clip- and frame-level annotations.
FineLAP integrates a dual-stream sigmoid loss with a cluster-based sampling strategy, enabling effective learning from data of varying granularities. 
To jointly model global semantics and local details, we design a decoupled audio adapter built on top of a self-supervised audio encoder, extracting dense frame-level representations and global embeddings in a unified framework. 
To further address data scarcity, we introduce \textbf{FineLAP-100k}, a large-scale synthetic SED dataset generated via a scalable pipeline. 
Leveraging the proposed training paradigm, model architecture, and dataset, FineLAP achieves SOTA performance across a wide range of audio understanding tasks, including retrieval, classification, sound event detection, and audio grounding. 
We further provide comprehensive ablation studies to validate our design components, offering insights for developing stronger audio-language models (ALMs).
Overall, our main contributions are as follows:
\begin{itemize}[itemsep=1.5pt, topsep=1.5pt, parsep=0pt, partopsep=0pt] 
\item We propose \textbf{FineLAP}, a novel training paradigm that exploits heterogeneous supervision via dual-stream sigmoid loss and cluster-based sampling to advance both coarse- and fine-grained alignment in CLAP.
\item We design a decoupled audio adapter to jointly capture global semantics and local details, enabling precise frame- and clip-level understanding.
\item We introduce \textbf{FineLAP-100k}, a large-scale synthetic SED dataset constructed via a scalable pipeline to mitigate data scarcity. 
\item FineLAP achieves SOTA performance across extensive audio understanding benchmarks. Our analysis further reveals that clip- and frame-level objectives are mutually reinforcing, paving the way for stronger ALMs. 
\end{itemize}

\noindent We will fully release the FineLAP codebase and model weights to facilitate future research.

\begin{figure}
    \centering
    \includegraphics[width=\linewidth]{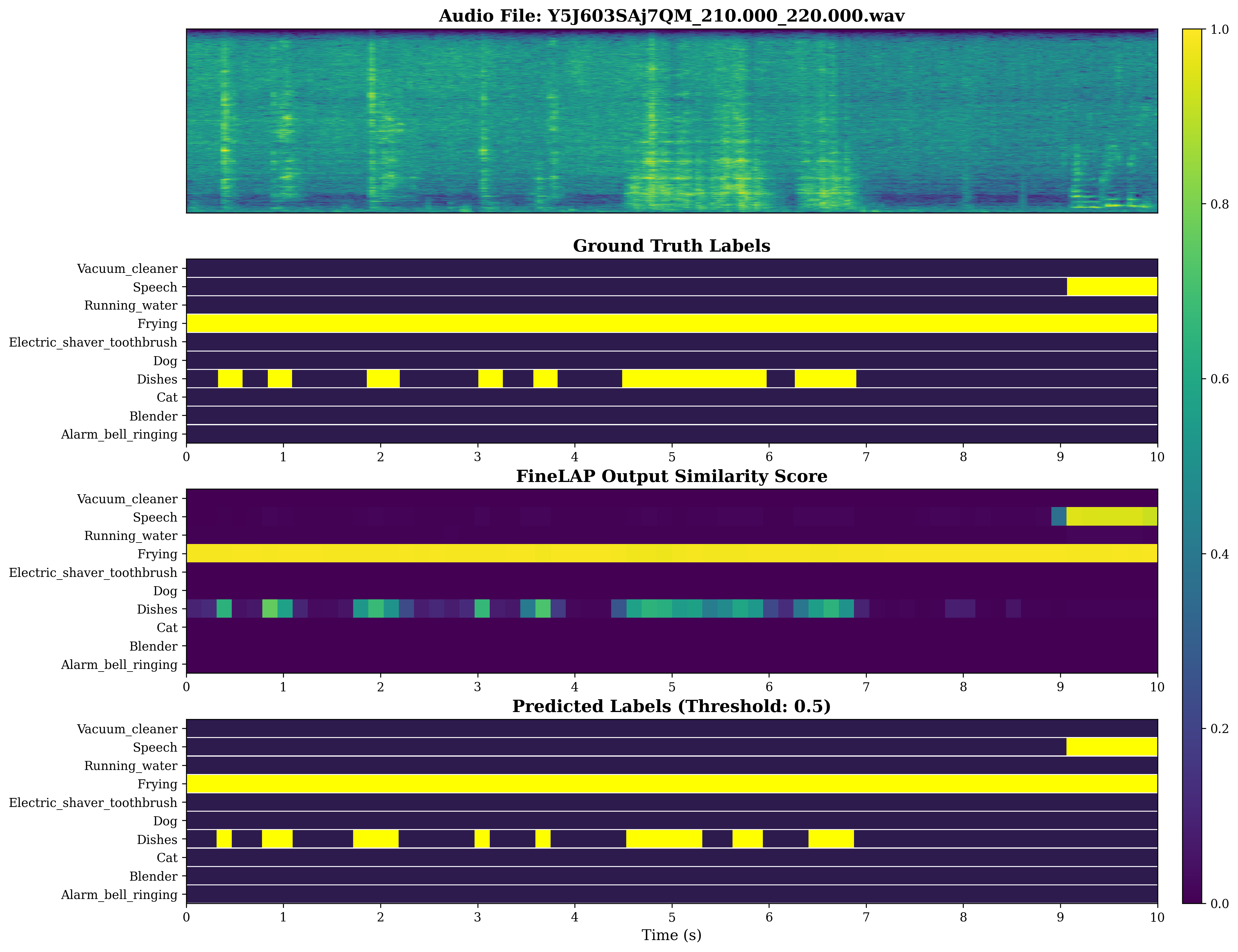}
    \caption{
    FineLAP effectively leverages the fine-grained alignment between audio frames and textual phrases to detect temporal boundaries of overlapping events. 
    }
    \label{fig:sed_results_demo}
    \vspace{-0.4cm}
\end{figure}

\begin{figure*}
    \centering
    \includegraphics[width=\linewidth]{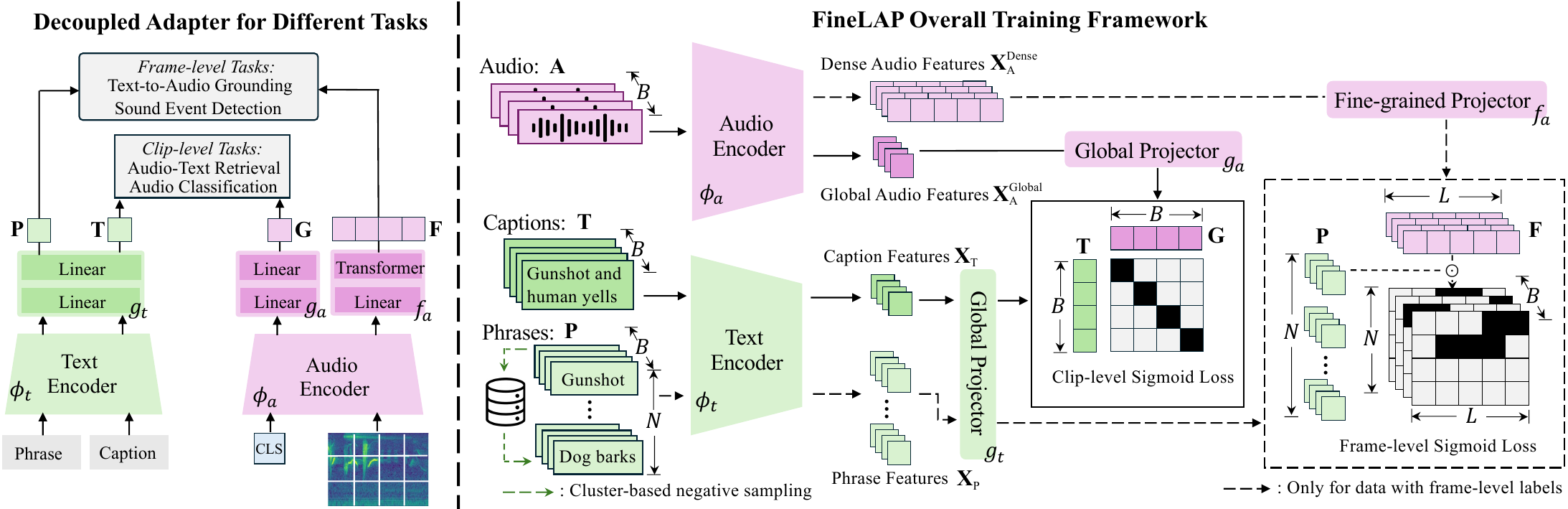}
    \caption{Overview of FineLAP. Left: The decoupled adapter enables simultaneous clip- and frame-level understanding. Right: The framework synergizes global and dense alignment by leveraging heterogeneous supervision via a dual-stream sigmoid loss with cluster-based sampling.
    }
    \label{fig:model_finelap}
    \vspace{-0.35cm}
\end{figure*}

\section{Preliminary: Contrastive Language Audio Pretraining}
\label{sec:preliminary_clap}
CLAP adopts a bi-encoder architecture to map audio and text into a shared multi-modal embedding space.
Given a batch of $B$ audio-caption pairs $\{(A^{(i)}, T^{(i)})\}_{i=1}^{B}$, the audio encoder $\phi_a$ and text encoder $\phi_t$ of CLAP first extract modality-specific representations:
$\mathbf{X}_A^{(i)} = \phi_a(A^{(i)})$, $\mathbf{X}_T^{(i)} = \phi_t(T^{(i)})$.
These representations are then projected by the global audio and text projectors into aligned clip- and sentence-level embeddings,
$\mathbf{A}_i = g_a(\mathbf{X}_A^{(i)}) \in \mathbb{R}^{1 \times d}$ and
$\mathbf{T}_i = g_t(\mathbf{X}_T^{(i)}) \in \mathbb{R}^{1 \times d}$.
Finally, the InfoNCE \cite{oord2018representation} loss is minimized to pull paired audio-text embeddings closer while pushing apart unpaired ones: 
\begin{equation*}
\begin{aligned}
&\mathcal{L}_{\text{InfoNCE}} = \\
& -\frac{1}{2B}\sum_{i=1}^{B}
\left(
\log \frac{e^{t \cdot s_{ii}}}{\sum_{j=1}^{B} e^{t \cdot s_{ij}}}
+
\log \frac{e^{t \cdot s_{ii}}}{\sum_{j=1}^{B} e^{t \cdot s_{ji}}}
\right)
\end{aligned}
\label{eq:infonce}
\end{equation*}
Where $s_{ij} = \frac{\mathbf{A}_i \cdot\ \mathbf{T}_j}{\| \mathbf{A}_i \| \cdot \| \mathbf{T}_j\|}$
denotes the cosine similarity between the $i$-th global audio embedding and the $j$-th caption embedding. 

\section{Fine-grained Language-Audio Pretraining}
\subsection{Notation}
We train FineLAP on heterogeneous data to jointly learn clip- and frame-level audio--language alignment.
Specifically, consider a training batch of $B$ audio-caption pairs
$\{(A^{(i)}, T^{(i)})\}_{i=1}^{B}$.
For some samples, the audio clip $A^{(i)}$ is additionally annotated with frame-level supervision, represented as
$
\{(P^{(i)}_k, Y^{(i)}_k)\}_{k=1}^{K_i},
$
where $\{P^{(i)}_k\}_{k=1}^{K_i}$ denotes a set of phrases describing brief acoustic events, and
$Y^{(i)}_k \in \{0,1\}^{L}$ is a frame-wise binary label over the $L$ frames of $A^{(i)}$.
For each phrase $P^{(i)}_k$, $Y^{(i)}_k[l] = 1$ indicates the presence of the corresponding event at frame $l$, and $0$ otherwise.
Since frame-level annotations are costly, most training samples provide only clip-level captions.
To sum up, we represent our training data as
\[
\left\{\left(A^{(i)}, T^{(i)}, \{(P^{(i)}_k, Y^{(i)}_k)\}_{k=1}^{K_i}\right)\right\}_{i=1}^{B},
\]
where the phrase set $\{(P^{(i)}_k, Y^{(i)}_k)\}$ is empty for samples without frame-level annotations, and only $(A^{(i)}, T^{(i)})$ is used during training.

\subsection{Model Architecture}
\paragraph{Audio Encoder.}
FineLAP employs EAT~\cite{chen2024eat} as its audio encoder $\phi_a(\cdot)$.
Trained with a self-supervised utterance-frame objective and an inverse block masking strategy, EAT achieves strong performance across a wide range of downstream audio tasks, consistently outperforming prior audio encoders such as HTS-AT~\cite{chen2022hts} and BEATs~\cite{chen2022beats}.

Given an audio clip $A$, EAT converts the raw waveform into a mel-spectrogram and partitions it into non-overlapping patches $\mathbf{X} \in \mathbb{R}^{L \times F \times d''}$, where $L$ and $F$ denote the numbers of temporal and frequency patches, respectively.
These patches, together with a newly added \texttt{[CLS]} token, are concatenated and fed into the backbone Transformer \cite{vaswani2017attention}. Finally, the model outputs a global clip-level representation $\mathbf{X}_{A}^{\mathrm{Global}} \in \mathbb{R}^{1 \times d'}$ corresponding to the \texttt{[CLS]} token, as well as dense frame-level audio features $\mathbf{X}_{A}^{\mathrm{Dense}} \in \mathbb{R}^{L \times F \times d'}$. 

\paragraph{Decoupled Audio Adapter.}
As illustrated in Figure \ref{fig:model_finelap} (left), FineLAP leverages both global and dense audio features for downstream tasks.
Since these two types of features capture fundamentally different semantic and temporal characteristics, we adopt a decoupled audio adapter architecture to process them separately.
Specifically, the global audio feature $\mathbf{X}_A^\text{Global}$ is projected into the shared audio-language embedding space by a global audio projector $g_a(\cdot)$, implemented as two linear layers, producing $\mathbf{G} \in \mathbb{R}^{1 \times d}$.
In parallel, the dense frame-level features $\mathbf{X}_A^\text{Dense}$ are first pooled along the frequency dimension and mapped into the shared embedding space via a linear projection.
Subsequently, two Transformer layers are applied to refine temporal dependencies, yielding $\mathbf{F} \in \mathbb{R}^{L \times d}$.
This decoupled asymmetric design accommodates the distinct requirements of clip-level semantic extraction and frame-level temporal modeling, as demonstrated by our experiments in Section \ref{sec:ablation_studies}.

\paragraph{Text Tower. }
We adopt RoBERTa~\cite{liu2019roberta} as the text encoder $\phi_t(\cdot)$ in FineLAP.
Unlike the audio branch which requires representations at multiple temporal granularities, the text branch focuses on global semantic understanding and does not involve localized modeling.
Accordingly, we employ a lightweight global text projector $g_t(\cdot)$, implemented as two linear layers, to project the holistic textual features of captions and phrases produced by RoBERTa.
For a given caption $T$ or phrase $P$, the text tower outputs
$\mathbf{T} = g_t(\phi_t(T)) \in \mathbb{R}^{1 \times d'}$
and
$\mathbf{P} = g_t(\phi_t(P)) \in \mathbb{R}^{1 \times d'}$,
respectively.

\subsection{Training Objective}

\paragraph{Clip-level Sigmoid Loss.}
Given a training batch of $B$ samples $\{(A^{(i)}, T^{(i)}, \{(P^{(i)}_k, Y^{(i)}_k)\}_{k=1}^{K_i})\}_{i=1}^{B}$,
we adopt a clip-level sigmoid loss \cite{zhai2023sigmoid} to align each audio clip with its corresponding caption.
Specifically, let $\mathbf{G}^{(i)} = g_a(\phi_a(A^{(i)}))$ and
$\mathbf{T}^{(i)} = g_t(\phi_t(T^{(i)}))$
denote the global audio and caption embeddings, respectively.
We minimize: 
\[
\mathcal{L}_\text{Global}
=
-\frac{1}{B}\sum_{i=1}^{B}\sum_{j=1}^{B}
\log \frac{1}{1 + e^{z_{ij}(-t \cdot s_{ij} + b)}} ,
\]
Here, $s_{ij} = \frac{\mathbf{G}^{(i)} \cdot \mathbf{T}^{(j)}} {\|\mathbf{G}^{(i)}\| \cdot \|\mathbf{T}^{(j)}\|}$ denotes the cosine similarity between the $i$-th global audio embedding and the $j$-th caption embedding. The binary label $z_{ij} = 1$ if $(A^{(i)}, T^{(j)})$ is a matched audio-caption pair and $z_{ij} = -1$ otherwise. The parameters $t$ and $b$ are learnable temperature and bias terms, respectively.
Unlike the InfoNCE loss used in standard CLAP models, the sigmoid loss operates on independent audio-text pairs and does not require computing global normalization factors, providing a consistent formulation with our frame-level objective. 
Empirical results in Appendix~\ref{app:additional_experiments} further confirm that this design leads to more stable optimization and improved overall performance.

\paragraph{Frame-level Sigmoid Loss.}
While clip-level supervision aligns an entire audio clip with its textual description, it does not provide explicit temporal grounding for individual phrases.
To enable fine-grained frame-phrase alignment, we introduce an additional frame-level sigmoid loss with cluster-based negative sampling.

Specifically, before training, we gather all phrases from the training corpus to form a database $\mathcal{D}_P$.
We then map each phrase $P \in \mathcal{D}_P$ into a set of pre-defined semantic clusters $\mathcal{C}$ and define a phrase-to-cluster mapping $\psi(\cdot)$, where $\psi(P) \in \mathcal{C}$ denotes the cluster assignment of phrase $P$.
During training, for a given audio clip $A^{(i)}$ with an associated set of positive phrases $\{P^{(i)}_k\}_{k=1}^{K_i}$, we first identify the set of positive clusters induced by these phrases: $\mathcal{C}_{+}^{(i)} = \bigcup_{k=1}^{K_i} \{\psi(P^{(i)}_k)\}$.
Negative clusters are defined as $\mathcal{C} \setminus \mathcal{C}_{+}^{(i)}$, which induces a candidate negative phrase pool
$\mathcal{N}(A^{(i)}) = \{P \in \mathcal{D}_P \mid \psi(P) \notin \mathcal{C}_{+}^{(i)}\}$.
We then sample $N - K_i$ negative phrases from $\mathcal{N}(A^{(i)})$ and construct the fixed-size frame-level annotation set
$\{(P^{(i)}_k, Y^{(i)}_k)\}_{k=1}^{N}$ by combining all labeled positives with the sampled negatives.
Note that for negative phrases, the corresponding frame-level labels are set to zero for all frames, i.e., $Y^{(i)}_k[l] = 0$ for all frame indices $l$.
The complete sampling procedure is summarized in Algorithm~\ref{alg:cluster_negative_sampling_fixedN} of the Appendix.
By adopting cluster-based negative phrase sampling, we effectively mitigate the scarcity of phrase annotations and ensure that each training sample is paired with a sufficient number of true negative phrases.
In addition, the diversity and variability within each cluster encourage the model to learn more generalizable representations, making it better suited for open-vocabulary sound event detection.

Given the enriched annotation set $\{(P^{(i)}_k, Y^{(i)}_k)\}_{k=1}^{N}$, we can then get the corresponding phrase embeddings $\mathbf{P}^{(i)}_k = g_t(\phi_t(P^{(i)}_k))$ and the dense audio embeddings $\mathbf{F}^{(i)} = f_a(\phi_a(A^{(i)}))$, the frame-level sigmoid loss is defined as
\begin{equation*}
\begin{aligned}
&\mathcal{L}_\text{Local}
= \\
&-\frac{1}{BNL}\sum_{i=1}^{B}\sum_{k=1}^{N}\sum_{l=1}^{L}
\log \frac{1}{1 + e^{z_{ikl}(-t' \cdot s_{ikl} + b')}} 
\end{aligned}
\end{equation*}
Here, 
$s_{ikl}$ denotes the cosine similarity between the $l$-th frame embedding of audio $A^{(i)}$ and the $k$-th phrase embedding associated with $A^{(i)}$.
The label $z_{ikl} \in \{1, -1\}$ is determined by the frame-level annotation $Y^{(i)}_k$, where $z_{ikl} = 1$ if $Y^{(i)}_k[l] = 1$ and $z_{ikl} = -1$ otherwise.
The parameters $t'$ and $b'$ are learnable temperature and bias terms, respectively.

To sum up, the combined objective
\[
\mathcal{L}_\text{Total} = \mathcal{L}_\text{Global} + \mathcal{L}_\text{Local}
\]
promotes both coarse- and fine-grained alignment with heterogeneous data. 
Our experiments further demonstrate that these two losses can reinforce each other, allowing the model to better exploit heterogeneous supervision for both clip- and frame-level learning.

\section{FineLAP-100k: A Large-scale Synthetic SED Dataset}
\subsection{Overview}
Due to the prohibitive cost of human labeling, high-quality SED data with frame-level annotations remain scarce. This challenge is particularly pronounced in open-vocabulary tasks, where the long-tail nature of sound events exacerbates annotation difficulty. Motivated by this, and inspired by \cite{wu2025flam}, we propose a scalable simulation pipeline to construct \textbf{FineLAP-100k}, a large-scale synthetic SED dataset.

\subsection{Data Creation Pipeline}
\label{sec:data_curation_pipleine}

\paragraph{Audio Event Clipping.}
A prerequisite for synthesizing high-quality SED data is the availability of clean, single-source audio clips. 
Particularly, segments containing overlapping events or vague boundaries introduce ambiguity, making precise temporal annotation unreliable.
To this end, we propose a window-based audio clipping strategy to extract single-event audio segments from high-quality audio tagging datasets. Specifically, we adopt FSD50K~\cite{fonseca2021fsd50k}, a human-labeled dataset comprising 51,197 high-quality audio clips sourced from Freesound\footnote{\url{https://freesound.org/}}.
We first select audio clips annotated with a single event label. 
For each selected audio clip, we compute its energy envelope and apply a sliding window across the signal.
At each window position, we evaluate the mean energy within the window and compare it against a predefined threshold ($-20 \mathrm{dB}$) to identify non-silent regions. 
We then merge consecutive windows exceeding the threshold and extract the first continuous high-energy segment as a clean, single-source audio clip.
We retain clipped segments whose durations fall between 1s and 7.5s, resulting in a total of 19,775 high-quality single-event audio clips. 
Visualizations of our strategy are provided in Appendix~\ref{app:visualization_window-based_clipping}.



\paragraph{Background Audio Selection.}
Background ambiance without salient audio events is also crucial for synthesizing realistic data and improving robustness. 
We select recordings labeled \emph{Ambiance} from the Adobe Audition SFX Library (ASFX)~\cite{wilkins2023bridging}.
This dataset offers high-quality environmental recordings spanning diverse acoustic scenes (e.g., streets, public spaces). We segment the downloaded audio into non-overlapping $10\,\mathrm{s}$ clips, resulting in 1,765 background segments.

\paragraph{Random Mixing.}

After obtaining high-quality foreground and background audio, we perform random mixing to synthesize the final SED dataset. Specifically, we first sample one background audio clip and then randomly select $M \sim \mathcal{U}(1, 5)$ foreground audio events from the event set.
For each foreground event, if its duration is shorter than $3\,\mathrm{s}$, we randomly repeat it $R \sim \mathcal{U}(1, 3)$ times.
The selected foreground events are placed at random temporal positions along a $10\,\mathrm{s}$ timeline.
During mixing, for each foreground event, we independently sample a signal-to-noise ratio
$\mathrm{SNR} \sim \mathcal{U}(12\,\mathrm{dB}, 20\,\mathrm{dB})$
to control the relative energy between the foreground event and the background audio.
Then, the RMS amplitude of each foreground event is scaled with respect to the background audio according to its sampled SNR, after which the foreground and background signals are mixed.
After audio synthesis, we generate a clip-level caption by aggregating all selected events using a rule-based strategy. Specifically, we prompt a large language model (LLM) to produce diverse caption templates (e.g., ``This audio contains the sounds of \{\}.'') and fill them with the concatenated event phrases to obtain the final caption.

\begin{table}[t]
\centering
\resizebox{\linewidth}{!}{
\begin{tabular}{l c c c}
\toprule
\textbf{Methods}
& \textbf{Type}
& \textbf{Data$^\dag$}
& \textbf{Supervision$^\ddag$}
\\
\midrule
LAION-CLAP \cite{wu2023large} & CLAP & 2.7M & Global \\
HTSAT-BERT \cite{mei2024wavcaps} & CLAP & 0.5M & Global \\
Cacophony \cite{zhu2024cacophony} & CLAP & 4.1M & Global\\
M2D-CLAP \cite{niizumi2025m2d} & CLAP & 2.6M & Global\\
MGA-CLAP \cite{li2024advancing} & CLAP & 0.5M & Global \\
FLAM \cite{wu2025flam} & CLAP & 2.2M & Global + Frame \\
PE$_\text{A-Frame}$ \cite{vyas2025pushing} & CLAP & $O$(100M) & Global + Frame \\ 
FlexSED \cite{hai2025flexsed} & OpenSED& 0.1M & Frame \\
PT-SED \cite{schmid2024effective} & CloseSED  & 0.1M & Frame \\
\midrule 
\rowcolor{blue!10} FineLAP (Ours) & CLAP & 2.2M & Global + Frame \\
\bottomrule
\end{tabular}
}
\caption{
Overview of baseline models. 
$\dag$: Data denotes number of audio clips used for training. 
$\ddag$: Supervision indicates whether training relies on clip-level annotations (Global) or frame-level annotations (Frame).
}
\label{tab:baselines}
\vspace{-0.5cm}
\end{table}

\begin{table*}[t]
\centering
\resizebox{\textwidth}{!}{
\begin{tabular}{l c ccc ccc ccc ccc}
\toprule
\multirow{4}{*}{\textbf{Methods}} 
& \multicolumn{6}{c}{\textbf{AudioCaps Eval (\%)}}
& 
& \multicolumn{6}{c}{\textbf{Clotho Eval (\%)}} \\
\cmidrule(lr){2-7} \cmidrule(lr){9-14}

& \multicolumn{3}{c}{T--A Retrieval}
& \multicolumn{3}{c}{A--T Retrieval}
& 
& \multicolumn{3}{c}{T--A Retrieval}
& \multicolumn{3}{c}{A--T Retrieval} \\
\cmidrule(lr){2-4} \cmidrule(lr){5-7} \cmidrule(lr){9-11} \cmidrule(lr){12-14}

& R@1 & R@5 & R@10
& R@1 & R@5 & R@10
& 
& R@1 & R@5 & R@10
& R@1 & R@5 & R@10 \\
\midrule

LAION-CLAP \cite{wu2023large} & 35.1 & 71.9 & 83.7 & 44.2 & 80.8 & 90.3 &  & 16.9 & 41.6 & 54.4 & 24.4 & 49.3 & 65.7 \\

HTSAT-BERT \cite{mei2024wavcaps}
& 39.7 & 74.5 & 86.1
& 51.7 & 82.3 & 90.6
&
& 19.5 & \underline{45.2} & 58.2
& 23.4 & 50.9 & 63.4 \\

Cacophony \cite{zhu2024cacophony}
& 41.0 & 75.3 & 86.4
& 55.3 & 83.6 & 92.4
&
& \underline{20.2} & \textbf{45.9} & \underline{58.8}
& \underline{26.5} & \textbf{54.1} & \textbf{67.3} \\

MGA-CLAP \cite{li2024advancing}
& \underline{42.2} & 74.9 & -
& 53.7 & 84.3 & -
&
& \textbf{20.8} & 45.0 & -
& \underline{26.5} & \textbf{54.1} & - \\

FLAM \cite{wu2025flam} & 32.1 & 64.8 & - & 43.3 & 75.0 & - & & 13.8 & 33.2 & - & 16.7 & 42.2 & -\\
M2D-CLAP \cite{niizumi2025m2d}
& 41.9 & \underline{77.1} & \underline{88.5}
& \underline{59.2} & \underline{84.7} & \underline{92.7}
&
& 20.1 & \textbf{45.9} & \textbf{59.4}
& 24.9 & 51.6 & 64.5 \\

\midrule

\rowcolor{blue!10}
FineLAP (Ours)
& \textbf{45.7} & \textbf{79.5} & \textbf{90.0}
& \textbf{62.5} & \textbf{85.9} & \textbf{95.0}
& & 18.9 & 43.5 & 56.8
& \textbf{26.6} & \underline{52.3} & \underline{66.0} \\

\bottomrule
\end{tabular}}
\caption{Audio-text retrieval performance on the evaluation split of AudioCaps and Clotho. 
}
\vspace{-0.3cm}
\label{tab:retrieval}
\end{table*}

\begin{table}[t]
\centering
\resizebox{\linewidth}{!}{
\begin{tabular}{l c c c c}
\toprule
\textbf{Methods}
& \textbf{DESED}
& \textbf{AS-SL}
& \textbf{USED}
& \textbf{TAG}
\\
\midrule

HTSAT-BERT \cite{mei2024wavcaps} & 0.131 & 0.284 & 0.016 & 0.344 \\
MGA-CLAP \cite{li2024advancing}
& 0.264 & 0.354 & 0.087 & 0.487
\\
FLAM \cite{wu2025flam} 
& 0.094 & 0.351 & 0.295 & -
\\
FlexSED \cite{hai2025flexsed} 
& 0.161 & 0.448 & 0.052 & -
\\
PE$_\text{A-Frame}$ \cite{vyas2025pushing} 
& \textbf{0.344} & 0.431 & 0.123 & -
\\
PT-SED \cite{schmid2024effective} 
& - & 0.465 & - & -
\\
\midrule
\rowcolor{blue!10}
FineLAP (Ours)
& \textbf{0.344} & \textbf{0.474} & \textbf{0.446} & \textbf{0.649}
\\

\bottomrule
\end{tabular}
}
\caption{Sound event detection performance on DESED, AudioSet-Strong (AS-SL), UrbanSED (USED) and AudioGrounding (TAG).
}
\label{tab:sed_results}
\end{table}

\begin{table}[t]
\centering
\resizebox{\linewidth}{!}{
\begin{tabular}{l c c c}
\toprule
\textbf{Methods}
& \textbf{ESC-50}
& \textbf{US8K}
& \textbf{VGGSound}
\\
\midrule
LAION-CLAP \cite{wu2023large} & 91.0 & 77.0 & 29.1 \\
HTSAT-BERT \cite{mei2024wavcaps} & 94.8 & 80.6 & 29.6 \\
Cacophony \cite{zhu2024cacophony}
& 93.4 & 77.1 & 27.0
\\
MGA-CLAP \cite{li2024advancing}
& \textbf{94.9} & 83.7 & 31.8
\\
FLAM \cite{wu2025flam}
& 86.9 & 75.6 & \textbf{39.3}
\\
M2D-CLAP \cite{niizumi2025m2d} & 94.3 & 82.3 & - \\
\midrule
\rowcolor{blue!10}
FineLAP (Ours) & 93.9 & \textbf{84.9} & 31.5 \\
\bottomrule
\end{tabular}
}
\caption{Audio classification performance on ESC-50, UrbanSound8K, and VGGSound.}
\vspace{-0.5cm}
\label{tab:audio_classification}
\end{table}

\section{Experimental Setup}

\subsection{Baselines}
We compare FineLAP with SOTA CLAP-based methods, including LAION-CLAP \cite{wu2023large}, HTSAT-BERT \cite{mei2024wavcaps}, Cacophony \cite{zhu2024cacophony}, MGA-CLAP \cite{li2024advancing}, FLAM \cite{wu2025flam}, M2D-CLAP \cite{niizumi2025m2d}, and PE$_\text{A-Frame}$ \cite{vyas2025pushing}.
Among these methods, only FLAM, MGA-CLAP, and PE$_\text{A-Frame}$ are explicitly designed to support frame-level audio–text alignment, where FLAM and PE$_\text{A-Frame}$ are directly trained with frame-level supervision.
To ensure a more comprehensive comparison, we additionally compare against SOTA closed-vocabulary SED (CloseSED) systems, namely PretrainedSED (PT-SED)~\cite{schmid2024effective}, as well as SOTA open-vocabulary SED (OpenSED) system FlexSED~\cite{hai2025flexsed}.
For PT-SED, we report the performance of its best-performing model using the BEATs encoder.
An overall comparison of these baselines is provided in Table~\ref{tab:baselines}, outlining the amount of training data (i.e., the number of audio samples), model type, and whether frame-level supervision is employed. Detailed descriptions of these baselines are given in Appendix~\ref{app:baseline_descriptions}.

\subsection{Datasets}
\paragraph{Training Data.}
For audio-text data without frame-level labels, we collect several large-scale open-source audio caption datasets, including the training splits of AudioSetCaps ~\cite{bai2025audiosetcaps}, WavCaps~\cite{mei2024wavcaps}, AudioCaps~\cite{kim2019audiocaps}, and Clotho~\cite{drossos2020clotho}.
For frame-level annotated data, we use the training sets of AudioSet-Strong~\cite{hershey2021benefit}, DESED-Strong~\cite{serizel2020sound}, UrbanSED~\cite{salamon2017scaper}, as well as our proposed FineLAP-100k dataset.
In total, we collect 2.1M audio-caption pairs without frame-level annotations, along with 201k samples with strong temporal annotations. The detailed dataset descriptions are provided in Appendix~\ref{app:training_data_details}.

For the pre-computed clustering space, we initialize the cluster centroids using phrases from the AudioSet-Strong ontology, which provides a comprehensive taxonomy covering a wide range of everyday sound events. However, the AudioSet-Strong ontology lacks fine-grained musical descriptions, as all music-related sounds are grouped under a single \emph{music} category. To address this limitation, we manually incorporate detailed musical instrument labels from the AudioSet ontology \cite{gemmeke2017audio} and further prompt a large language model (LLM) to systematically refine and expand the category system, aiming to improve semantic diversity and consistency. 
After obtaining the final set of cluster centroids, we encode all event phrases from the training data using Sentence-BERT \cite{reimers2019sentence} and assign each phrase to its nearest cluster based on cosine similarity. In total, our clustering space consists of 494 clusters with over 600 alternative phrases, covering a wide range of sound events.

\paragraph{Evaluation Data.}
We evaluate FineLAP on both clip- and frame-level tasks.
For clip-level tasks, we assess model's retrieval and classification performance.
For retrieval, we employ the test splits of AudioCaps and Clotho, using Recall@1,5,10 (R@1,5,10). 
For classification, we adopt the template ``The sound of \{\}'' and evaluate on ESC-50~\cite{piczak2015esc}, UrbanSound8K~\cite{salamon2014dataset}, and VGGSound~\cite{chen2020vggsound}, reporting classification accuracy.

For frame-level tasks, we evaluate sound event detection on the evaluation set of AudioSet-Strong, DESED, and UrbanSound-SED using the PSDS1 metric.  
For text-to-audio grounding, we report performance on the evaluation split of AudioGrounding~\cite{xu2021text} using the PSDS2021 metric.
The detailed computation of these evaluation metrics is provided in Appendix~\ref{app:evaluation_details}.


\begin{table*}[t]
\centering
\resizebox{\linewidth}{!}{
\begin{tabular}{l cc cccc cc}
\toprule
\multirow{2}{*}{\textbf{Methods}}
& \multicolumn{2}{c}{\textbf{Audio-Text Retrieval}}
& \multicolumn{4}{c}{\textbf{Sound Event Detection}}
& \multicolumn{2}{c}{\textbf{Audio Classification}} \\
\cmidrule(lr){2-3} \cmidrule(lr){4-7} \cmidrule(lr){8-9}
& AC-T2A$^*$ & AC-A2T$^*$
& DESED & AS-SL & UrbanSED & TAG
& VGGSound & US8K \\
\midrule

\rowcolor{blue!10}
FineLAP (Ours)
& 45.7 & \textbf{62.5}
& \textbf{0.344} & \textbf{0.474} & 0.446 & \textbf{0.649}
& 31.5 & 84.9 \\

\quad w/o Frame-level Loss
& 44.4 & 58.3
& 0.021 & 0.191 & 0.000 & 0.303
& 31.4 & 82.7 \\

\quad w/o Clip-level Loss 
& 4.9 & 6.2 
& 0.322 & 0.451 & \textbf{0.462} & 0.480 & 11.3 & 58.6\\

\quad w/o Decoupled Projector
& 44.8 & 55.4
& 0.317 & 0.467 & 0.442 & 0.448
& \textbf{31.8} & \textbf{85.4} \\

\quad w/o Synthetic Data
& 45.4 & 61.6
& 0.324 & 0.468 & 0.154 & 0.589
& 31.3 & 78.0 \\

FineLAP-InfoNCE & \textbf{46.0} & 61.0 
& 0.342 & \textbf{0.474} & 0.445 & 0.629  & 31.5 & 80.4
\\
FineLAP-HTSAT
& 41.8 & 56.8
& 0.292 & 0.416 & 0.273 & 0.569
& 30.6 & 47.8 \\


\bottomrule
\end{tabular}
}
\caption{Ablation studies on loss design, model architecture and synthetic data.
$^*$: AC-T2A and AC-A2T denote the R@1 retrieval performance on the evaluation split of AudioCaps.}
\vspace{-0.3cm}
\label{tab:ablation_loss_data_decouple}
\end{table*}

\subsection{Implementation Details}
We train FineLAP for 10 epochs with a batch size of 1024. The learning rate is set to $5 \times 10^{-5}$, using a cosine annealing scheduler with 1,000 warm-up steps. The total number of phrases is set to $N=20$. The embedding dimension is set to $d = 1024$. 
The learnable temperature and bias parameters are initialized as $t = t' = 10$ and $b = b' = -10$, respectively. 
During training, we pad or crop every audio clip to 10s.

\section{Experimental Results}
\subsection{Main Results}

\paragraph{Audio-text Retrieval.}

As shown in Table~\ref{tab:retrieval}, FineLAP achieves state-of-the-art retrieval performance on AudioCaps, reaching R@1 scores of 45.7 (T2A) and 62.5 (A2T), outperforming prior CLAP-based methods by a clear margin. 
On Clotho, FineLAP achieves comparable performance, with R@1 of 18.9 (T2A) and 26.6 (A2T), on par with existing state-of-the-art approaches. 
We attribute the smaller gains on Clotho to differences in data characteristics: Clotho contains variable-length audio, whereas FineLAP is trained and evaluated on fixed-length 10\,s segments, leading to a mismatch that may limit retrieval performance. 
Extending FineLAP to support variable-length inputs is left as future work. 
Overall, these results demonstrate the effectiveness of FineLAP in learning global audio–text alignment, with strong performance on both AudioCaps and Clotho.

\paragraph{Sound Event Detection.}
As shown in Table \ref{tab:sed_results}, FineLAP achieves the best overall performance across all evaluated SED benchmarks, outperforming both open-vocabulary and closed-vocabulary baselines by a clear margin. In particular, FineLAP attains the highest scores on DESED (0.344), AudioSet-Strong (0.474), UrbanSED (0.446), and TAG (0.649), demonstrating strong generalization across diverse datasets with different label spaces and annotation schemes.

Compared to CLAP-based methods that rely solely on global supervision (e.g., HTSAT-BERT and MGA-CLAP), FineLAP shows substantial gains, highlighting the importance of incorporating frame-level supervision for temporal localization. While prior approaches such as FLAM and PE$_\text{A-Frame}$ incorporate frame-level objectives, their performance still lags behind specialized SED models.
In contrast, FineLAP consistently improves performance across all evaluated datasets, suggesting that our training paradigm effectively leverages heterogeneous audio–text data to learn transferable frame-level representations.
It is worth noting that FineLAP operates in an open-vocabulary setting, offering greater flexibility than traditional closed-vocabulary SED models. This capability is further validated by its strong performance on the TAG benchmark, where FineLAP can effectively ground previously unseen event phrases.

\paragraph{Audio Classification.}
As shown in Table~\ref{tab:audio_classification}, FineLAP achieves strong audio classification performance across three benchmarks. It attains the best accuracy on UrbanSound8K (84.9\%) and remains competitive on ESC-50 (93.9\%) and VGGSound (31.5\%). 
These results indicate that FineLAP’s learned representations generalize well to standard audio classification tasks.

\subsection{Ablation Studies}
\label{sec:ablation_studies}
We conduct a comprehensive ablation study to assess the contribution of each design component. 
Specifically, we focus on the following three research questions:
\textbf{Q1}: Does heterogeneous supervision benefit both clip- and frame-level learning?
\textbf{Q2}: Does the proposed architecture improve performance at both granularities?
\textbf{Q3}: How much does synthetic data improve model's performance?

\paragraph{Question 1.}
We first investigate whether clip-level and frame-level supervision provide mutually reinforcing signals during training. 
For this, we begin by removing the frame-level loss and train the model using only clip-level sigmoid supervision. 
As shown in Table~\ref{tab:ablation_loss_data_decouple}, this leads to a severe degradation on frame-level tasks, with DESED and UrbanSED dropping from 0.344 and 0.446 to 0.021 and 0.000, respectively. 
Notably, clip-level performance also declines (e.g., AudioCaps A2T: 62.5 → 58.3, UrbanSound8K: 84.9 → 82.7), suggesting that the absence of fine-grained alignment weakens global audio–text representations.

We then remove the clip-level loss and train the model using only frame-level annotated data. 
To ensure a fair comparison, we keep the total number of optimization steps comparable to the main setting. 
As shown in Table~\ref{tab:ablation_loss_data_decouple}, removing clip-level supervision causes a substantial drop in retrieval and classification performance (AC-T2A: 45.7 → 4.9, AC-A2T: 62.5 → 6.2), and degrades frame-level performance (DESED: 0.344 → 0.322, AS-SL: 0.474 → 0.451). 
These results indicate that clip-level supervision provides essential semantic guidance for temporal localization.
Overall, these findings confirm that clip-level and frame-level supervision are complementary: 
the former provides coarse semantic alignment, while the latter enables precise temporal modeling, and their combination improves performance at both granularities.

We further compare sigmoid loss with InfoNCE loss (FineLAP-InfoNCE). 
While both objectives achieve comparable performance on audio–text retrieval, sigmoid loss consistently yields better results on downstream tasks. 
In particular, it improves sound event detection (DESED: 0.344 vs. 0.342, TAG: 0.649 vs. 0.629), while maintaining similar performance on AS-SL and UrbanSED. 
A similar trend is observed for audio classification, where sigmoid loss significantly improves UrbanSound8K (84.9 vs. 80.4). 
These results suggest that sigmoid loss provides a more stable and less competitive supervision signal, better preserving fine-grained information for frame-level modeling.

\paragraph{Question 2.}
We next investigate the role of the decoupled projector and the audio encoder.
To this end, we remove the fine-grained adapter $f_a$ and instead employ a single global projector $g_a$ to extract both holistic and dense audio embeddings.
As shown in Table~\ref{tab:ablation_loss_data_decouple}, this modification leads to consistent performance degradation across both retrieval and SED tasks.
Specifically, AC-A2T drops from 62.5 to 55.4, while SED performance on UrbanSED and TAG decreases from 0.446 and 0.649 to 0.442 and 0.448, respectively.

In addition, we replace the self-supervised EAT encoder with HTS-AT (denoted as FineLAP-HTSAT) while keeping the rest of the framework unchanged.
This substitution results in consistent performance drops on both retrieval and SED benchmarks, e.g., AC-A2T R@1 decreases from 62.5 to 56.8 and DESED performance drops from 0.344 to 0.292.
Together, these results indicate that effective multi-granularity alignment requires both a strong audio encoder and a decoupled projection design that separates global and fine-grained representations.
Additional ablation studies on the architectural choices of the fine-grained audio adapter are provided in Appendix~\ref{app:additional_experiments}.

\paragraph{Question 3.}
Finally, we analyze the effect of synthetic data on model performance.
As shown in Table~\ref{tab:ablation_loss_data_decouple}, removing the synthetic data leads to consistent performance drops across all benchmarks.
For example, performance on UrbanSED decreases from 0.446 to 0.154, while TAG drops from 0.649 to 0.589, indicating that synthetic data provides useful supervision for modeling temporal details. 
At the same time, we observe relatively modest changes in datasets such as AudioSet-SL (0.474 to 0.468) and DESED (0.344 to 0.324). 
We hypothesize that this behavior stems from discrepancies between the generated data and the target datasets, including differences in label taxonomy, event granularity, and underlying audio source distributions.



\section{Conclusion}
In this paper, we present FineLAP, a novel paradigm that advances both coarse- and fine-grained audio–text alignment in CLAP using heterogeneous supervision. 
Through a dual-stream sigmoid loss, FineLAP effectively learns from both frame- and clip-level labeled data. 
By introducing a decoupled audio adapter, FineLAP aligns multi-granularity representations from self-supervised audio encoders into a shared embedding space. 
To further mitigate data scarcity, we construct FineLAP-100k, a large-scale SED dataset curated via a scalable pipeline. 
Building upon the proposed training objective, model architecture, and dataset, FineLAP achieves SOTA performance across both clip- and frame-level audio understanding tasks. 
Comprehensive ablation studies further validate the effectiveness of each design component, paving the way towards better audio-language models.

\newpage
\section*{Limitations}
Despite its effectiveness in multi-granularity audio–language alignment, FineLAP has several limitations.
First, FineLAP does not explicitly support variable-length or long-form audio modeling.
The temporal resolution and input length are largely constrained by the underlying audio encoder, which is primarily designed for short- to medium-length audio clips. While this design choice enables stable training and efficient scaling, it limits FineLAP’s applicability to long-form sound event detection scenarios, particularly those involving dense, overlapping events and long-range temporal dependencies.
Second, frame-level evaluation in this work is mainly focused on sound event detection.
Although SED serves as a foundational task for frame-level audio understanding, other temporally grounded audio–language tasks remain unexplored. In particular, tasks such as temporal question answering and temporally enhanced audio–text retrieval are not included in the current evaluation and represent important directions for future research.



\bibliography{custom}

\appendix

\begin{table*}[ht]
\centering
\resizebox{\linewidth}{!}{
\begin{tabular}{ll cc cccc cc}
\toprule
\multirow{2}{*}{\textbf{Methods}}
& \multirow{2}{*}{\makecell{\textbf{Fine-grained}\\\textbf{Projector}}}
& \multicolumn{2}{c}{\textbf{Audio-Text Retrieval}}
& \multicolumn{4}{c}{\textbf{Sound Event Detection}}
& \multicolumn{2}{c}{\textbf{Audio Classification}} \\
\cmidrule(lr){3-4} \cmidrule(lr){5-8} \cmidrule(lr){9-10}
& 
& AC-T2A & AC-A2T
& DESED & AS-SL & UrbanSED & TAG
& VGGSound & US8K \\
\midrule
\rowcolor{blue!10}
FineLAP (Ours)
& Transformer
& 45.7 & \textbf{62.5}
& \textbf{0.344} & \textbf{0.474} & \textbf{0.446} & \textbf{0.649}
& \textbf{31.5} & \textbf{84.9} \\
& BiGRU
& 45.8 & 61.7
& 0.329 & 0.471 & 0.432 & 0.589
& 31.1 & 84.5 \\
& Linear
& \textbf{46.3} & 61.1
& 0.329 & 0.463 & 0.435 & 0.570
& 31.5 & 83.0 \\
\bottomrule
\end{tabular}
}
\caption{Ablation studies on the architecture of fine-grained audio adapter $f_a$.}
\label{app_tab:ablation_fa_architecture}
\end{table*}



\section{Additional Ablation Studies}
\label{app:additional_experiments}
\paragraph{Architecture of the Fine-grained Audio Adapter.}
We provide additional ablation studies on the architectural choice of the fine-grained audio projector $f_a$.
As illustrated in Table~\ref{app_tab:ablation_fa_architecture}, we evaluate three alternative designs for $f_a$, including a Transformer-based projector, a two-layer Bidirectional GRU (BiGRU), and a two-layer linear projection.
Overall, the Transformer-based projector achieves the most consistent performance across both clip-level audio--text retrieval and frame-level sound event detection tasks. In particular, it yields clear improvements on frame-level benchmarks such as DESED, AudioSet-Strong, UrbanSED, and TAG, suggesting that self-attention is effective at modeling fine-grained temporal dependencies in frame-level alignment. While the BiGRU and linear projectors achieve comparable results on audio--text retrieval, their performance on frame-level SED tasks is consistently lower, indicating limited temporal modeling capacity.
Importantly, these findings support the asymmetric design of FineLAP, where a lightweight global adapter is paired with a more expressive fine-grained projector. Such a decoupled architecture allows FineLAP to preserve strong clip-level alignment while enhancing frame-level modeling capacity, resulting in effective multi-granularity audio-language learning.

\subsection{Sensitivity Analysis on Clustering Design}

We conduct a sensitivity analysis on the cluster-based negative sampling design (Algorithm \ref{alg:cluster_negative_sampling_fixedN}). 
Specifically, we vary $N$, the maximum number of phrases (including padded negative phrases) constructed for each audio sample. 
The model is trained on AudioSet-Strong and evaluated on DESED using the PSDS metric.

\begin{table}[ht]
\centering
\begin{tabular}{c|c}
\toprule
$N_{\text{phrases}}$ & DESED-PSDS $\uparrow$ \\
\midrule
10 & 0.163 \\
15 & 0.175 \\
20 & 0.182 \\
25 & \textbf{0.184} \\
30 & 0.181 \\
\bottomrule
\end{tabular}
\caption{Sensitivity analysis on the number of phrases $N$ in the clustering-based negative sampling design.}
\label{tab:ablation_cluster_N}
\end{table}

As shown in Table~\ref{tab:ablation_cluster_N}, increasing $N$ from 10 to 25 steadily improves performance, indicating that incorporating more diverse negative phrases enhances the model’s discriminative ability for fine-grained sound events. 
However, performance saturates and slightly degrades when $N$ increases to 30, suggesting that excessive negative phrases may introduce noise or weaken the learning signal. 
Based on this observation, we adopt $N = 20$ in our main experiments, which achieves a favorable balance between performance and computational efficiency.



\section{Visualizations}
\subsection{SED Results Visualization}
\label{app:sed_viz}
In this section, we present qualitative visualizations of FineLAP's performance on the sound event detection (SED) task. 
We conduct evaluations on the test sets of AudioSet-Strong and DESED. 
For all visualized samples, the final predicted labels are obtained by applying a fixed threshold of $0.5$ to the output probabilities. 
The specific visualization results are provided below.

\begin{figure*}
    \centering
    \includegraphics[width=0.49\linewidth]{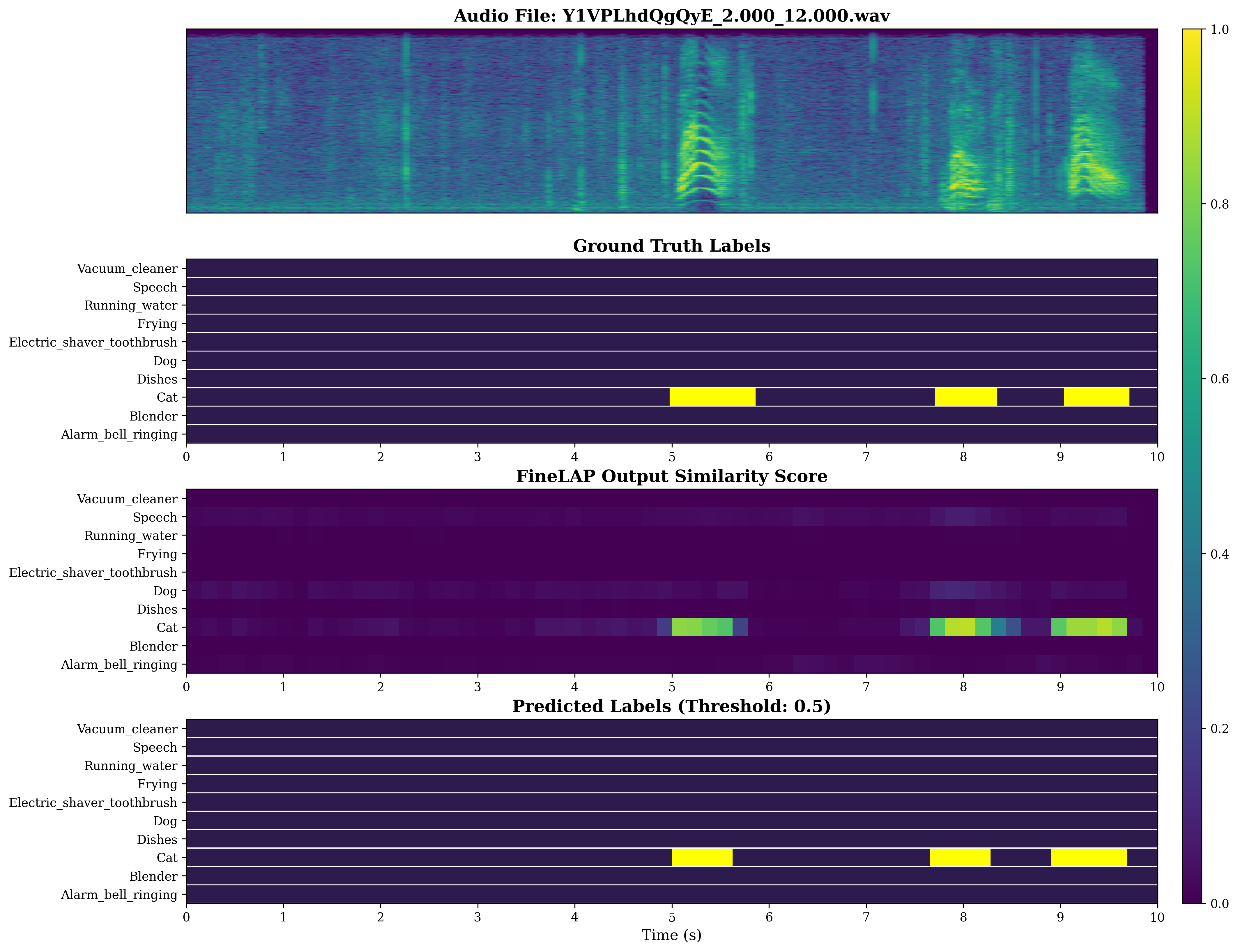} \hfill
    \includegraphics[width=0.49\linewidth]{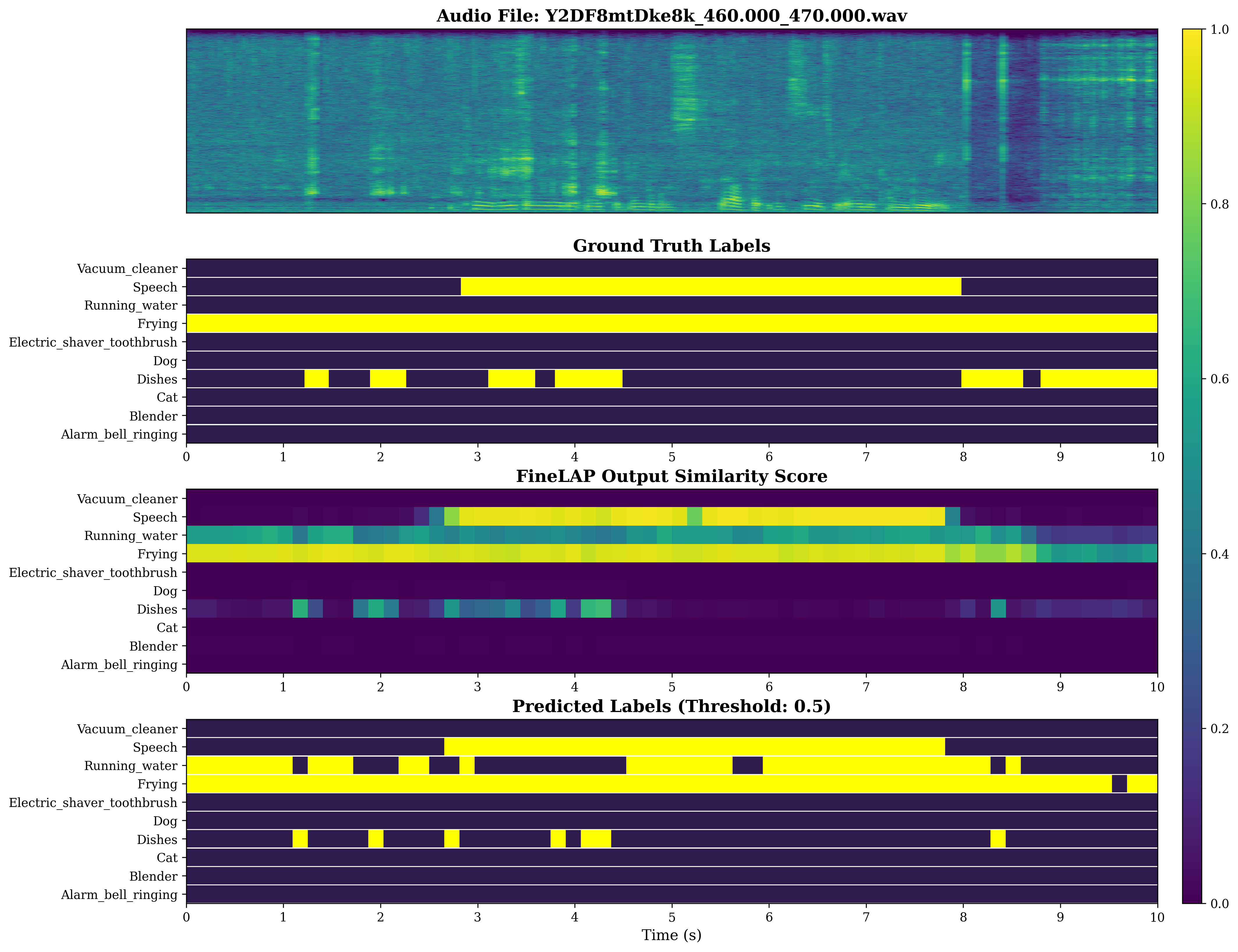}
      \vspace{0.5cm}
    \includegraphics[width=0.49\linewidth]{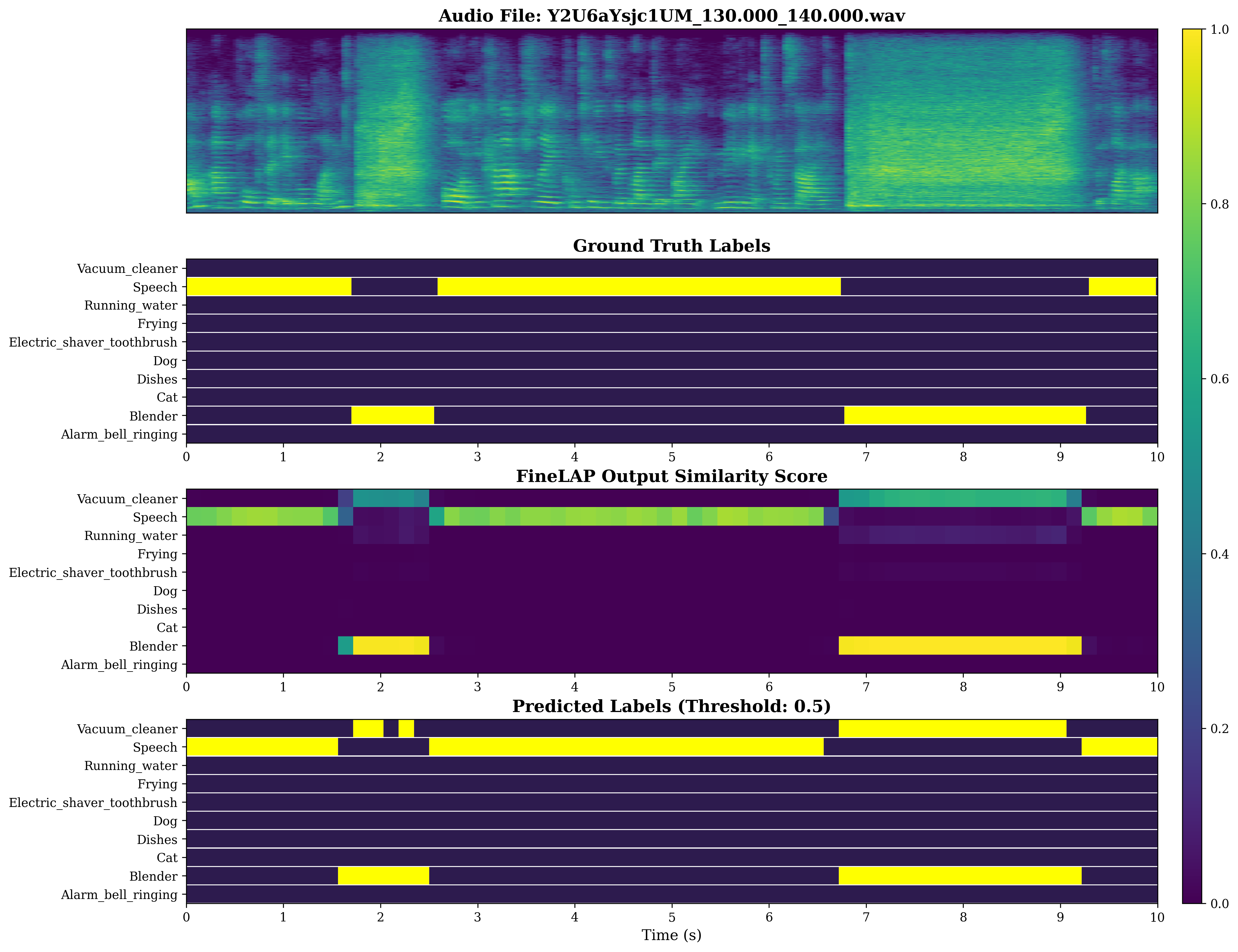} \hfill
     \includegraphics[width=0.49\linewidth]{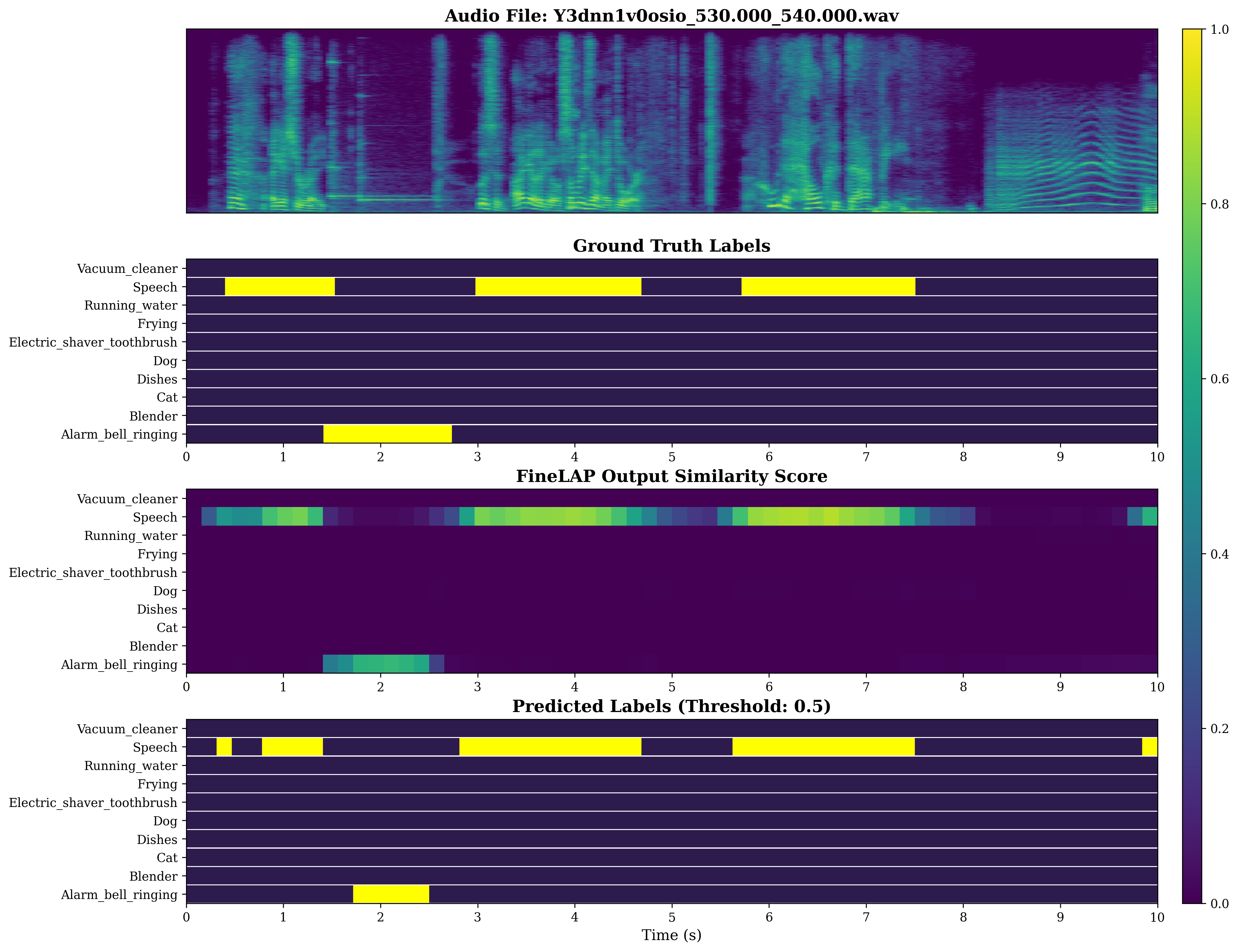}
    \vspace{0.5cm}
    \includegraphics[width=0.49\linewidth]{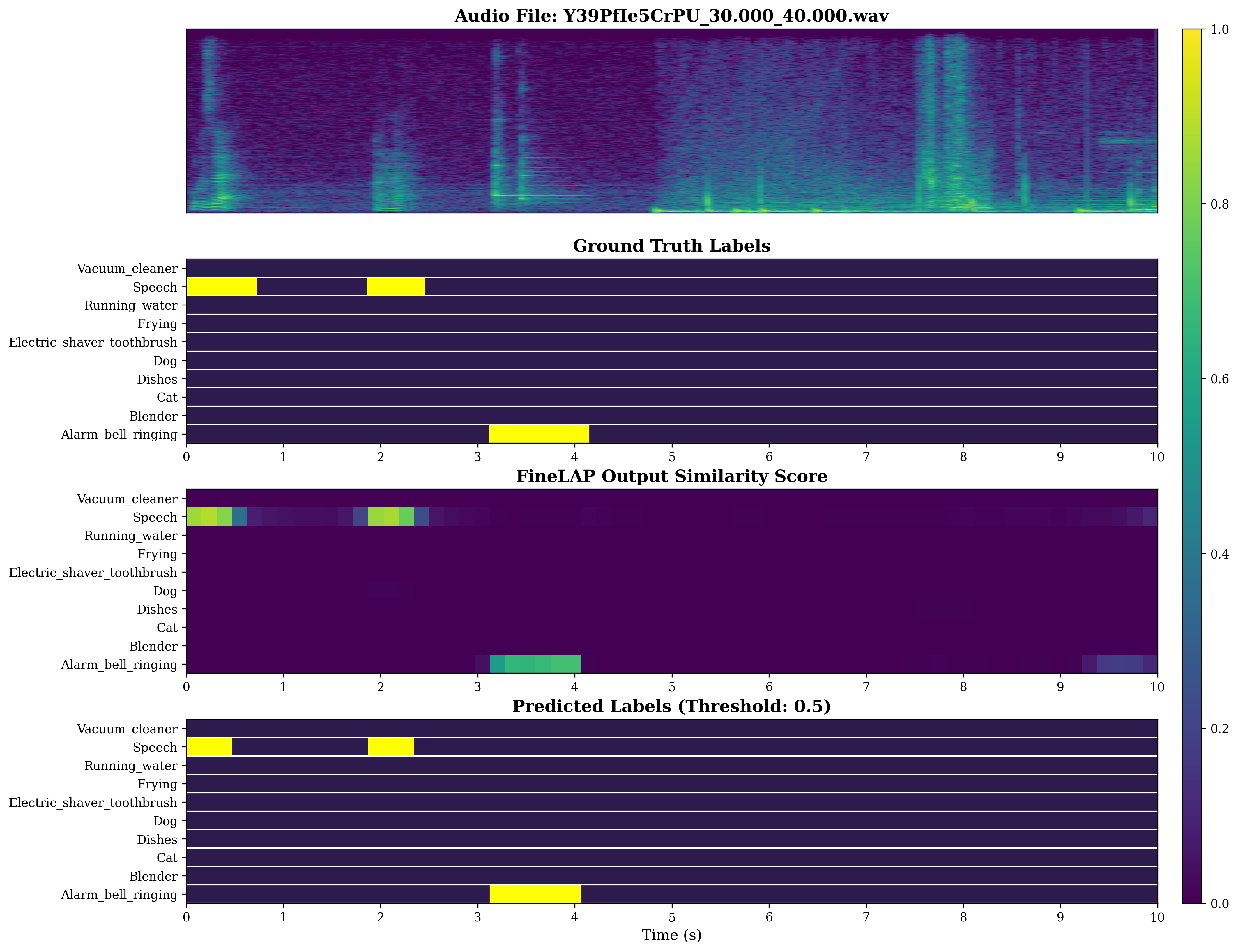} \hfill
     \includegraphics[width=0.49\linewidth]{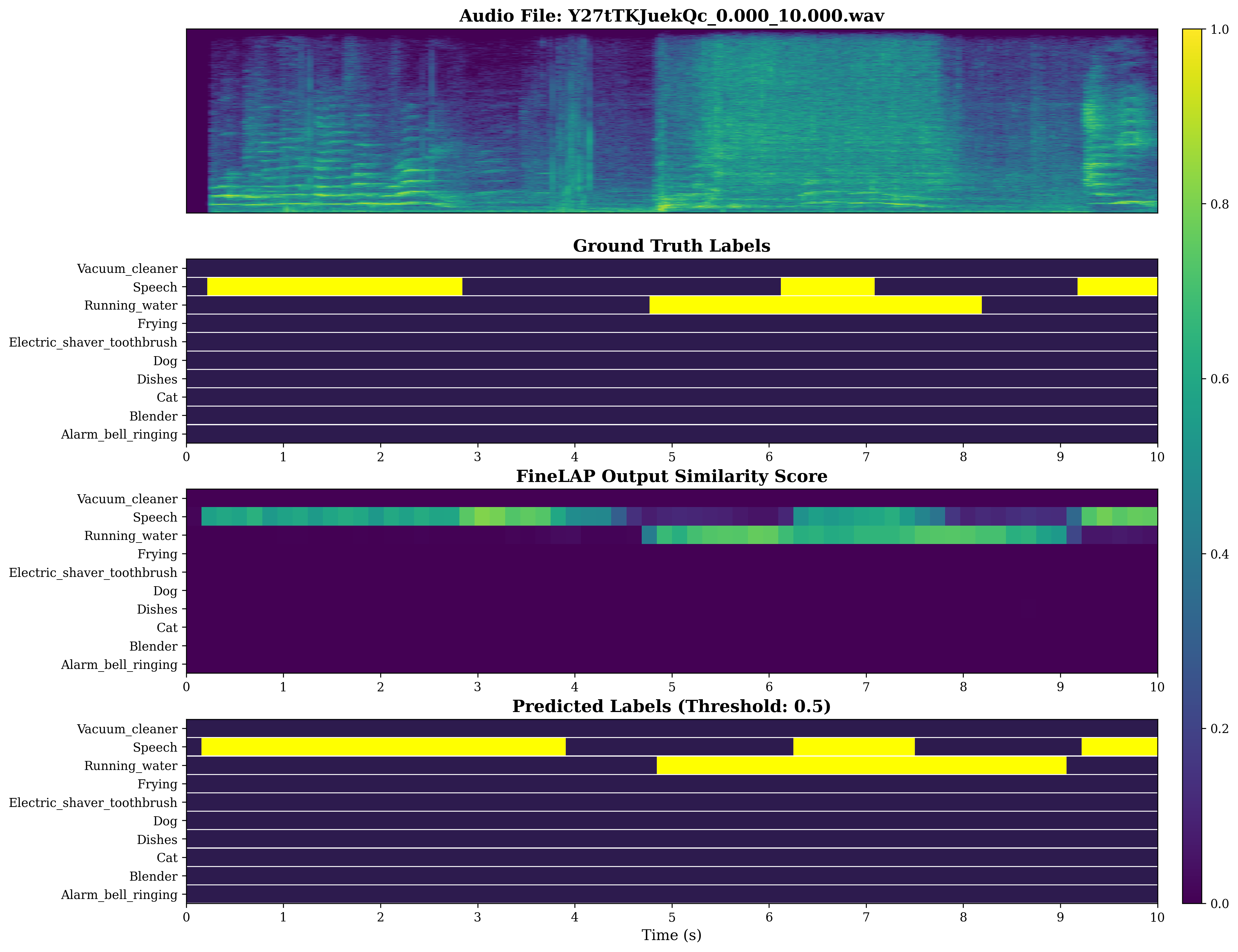}
    \caption{Visualization of FineLAP's performance on DESED-Eval }
\end{figure*}

\begin{figure*}
    \centering
    \includegraphics[width=0.49\linewidth]{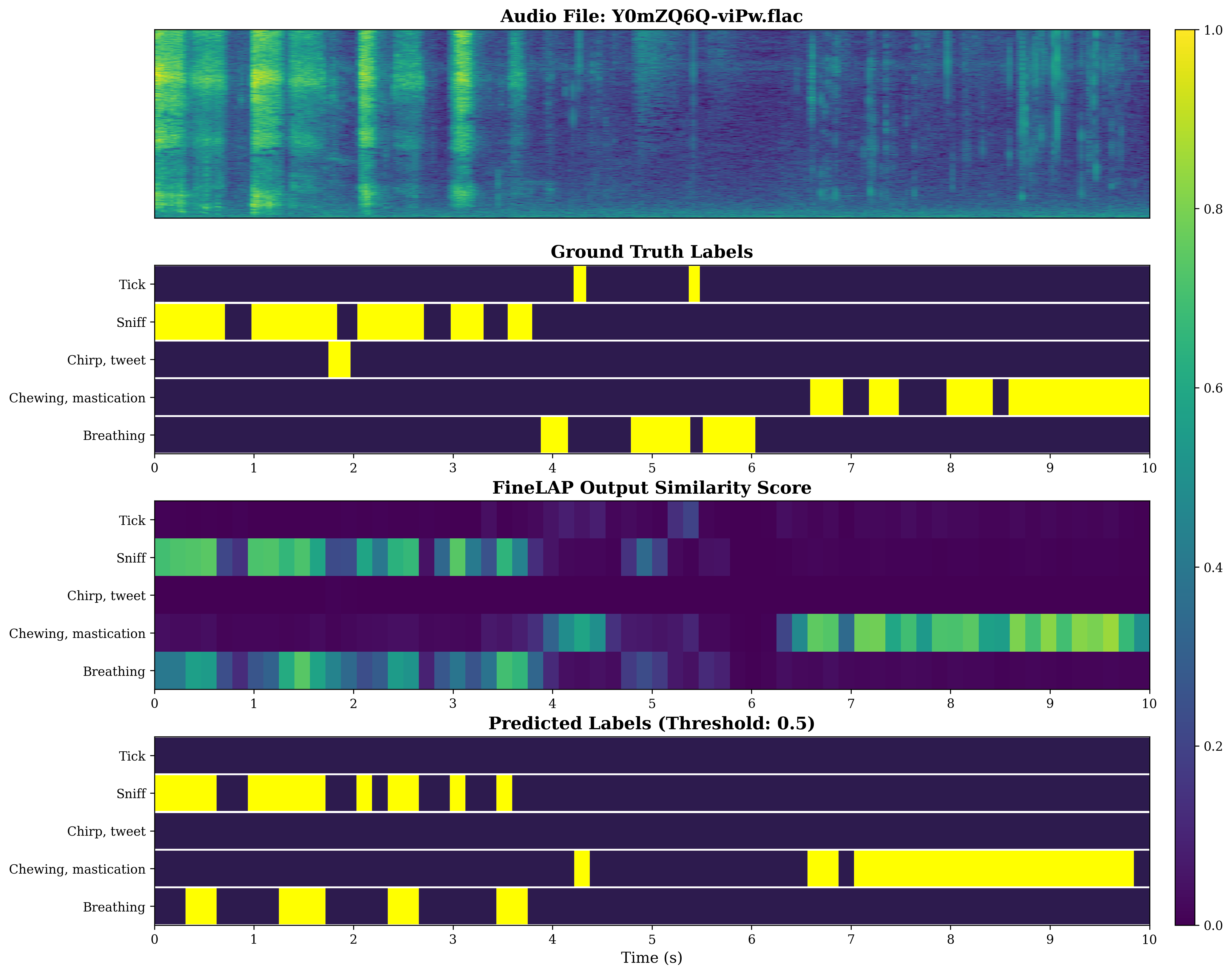} \hfill
    \includegraphics[width=0.49\linewidth]{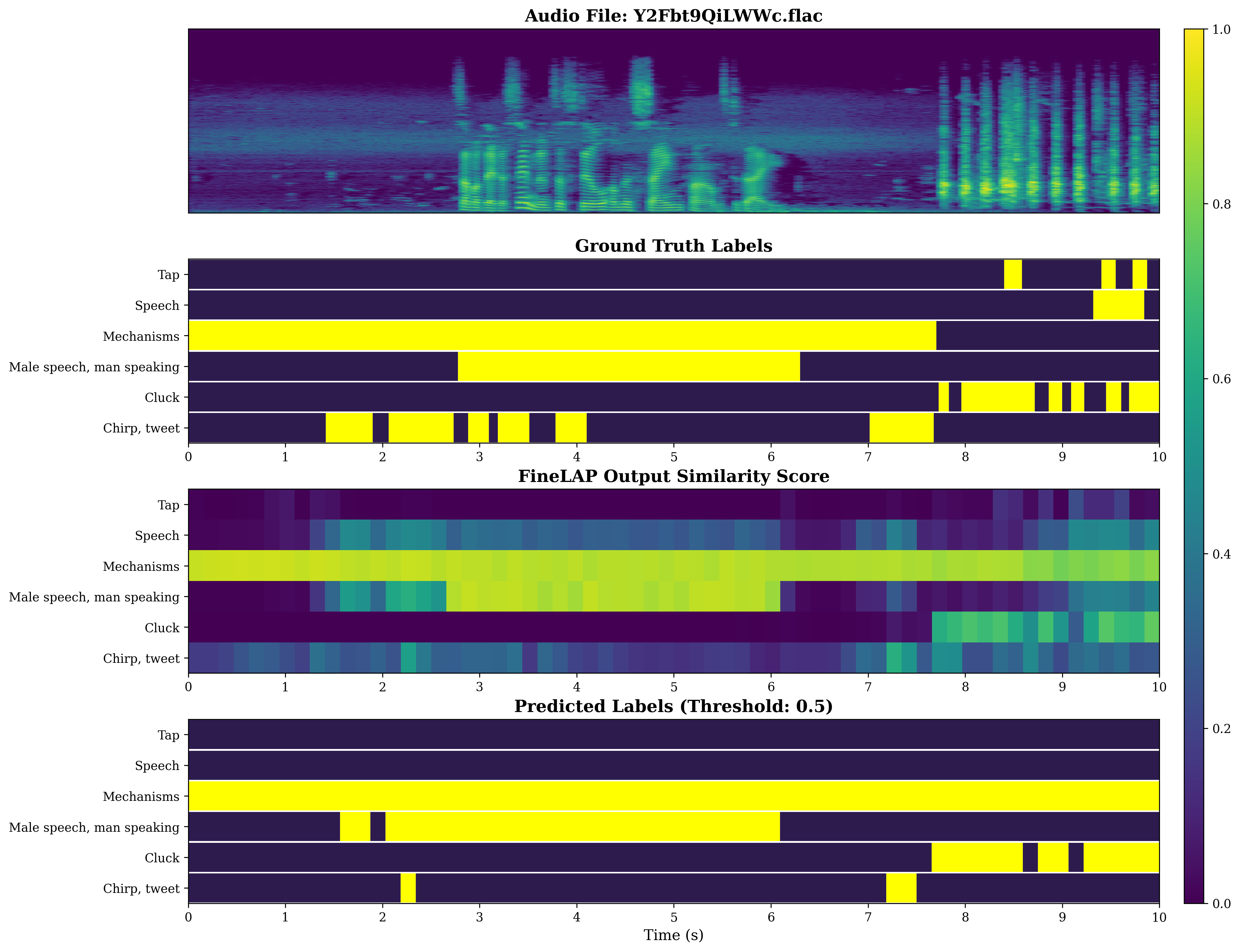}
      \vspace{0.5cm}
    \includegraphics[width=0.49\linewidth]{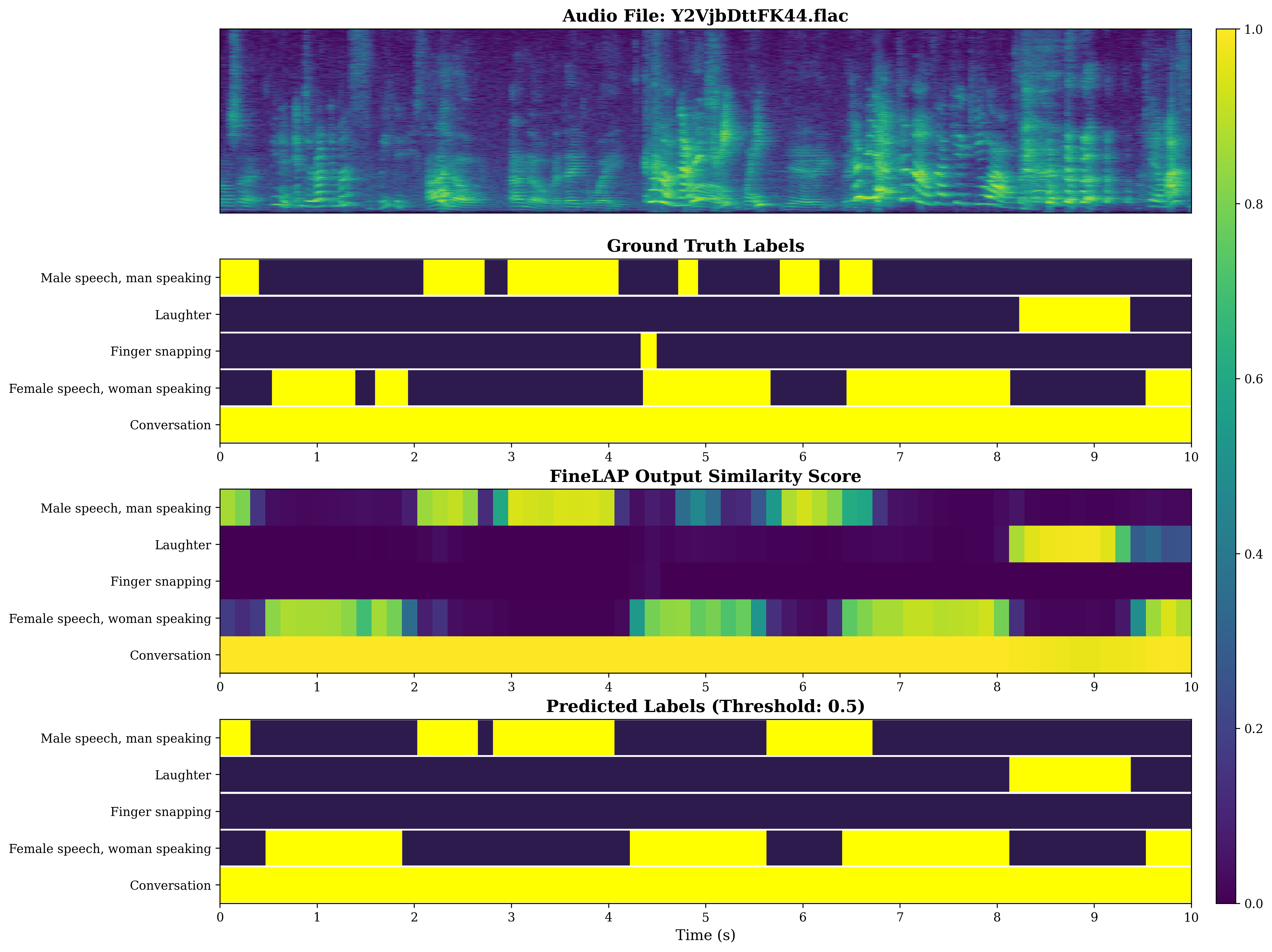} \hfill
     \includegraphics[width=0.49\linewidth]{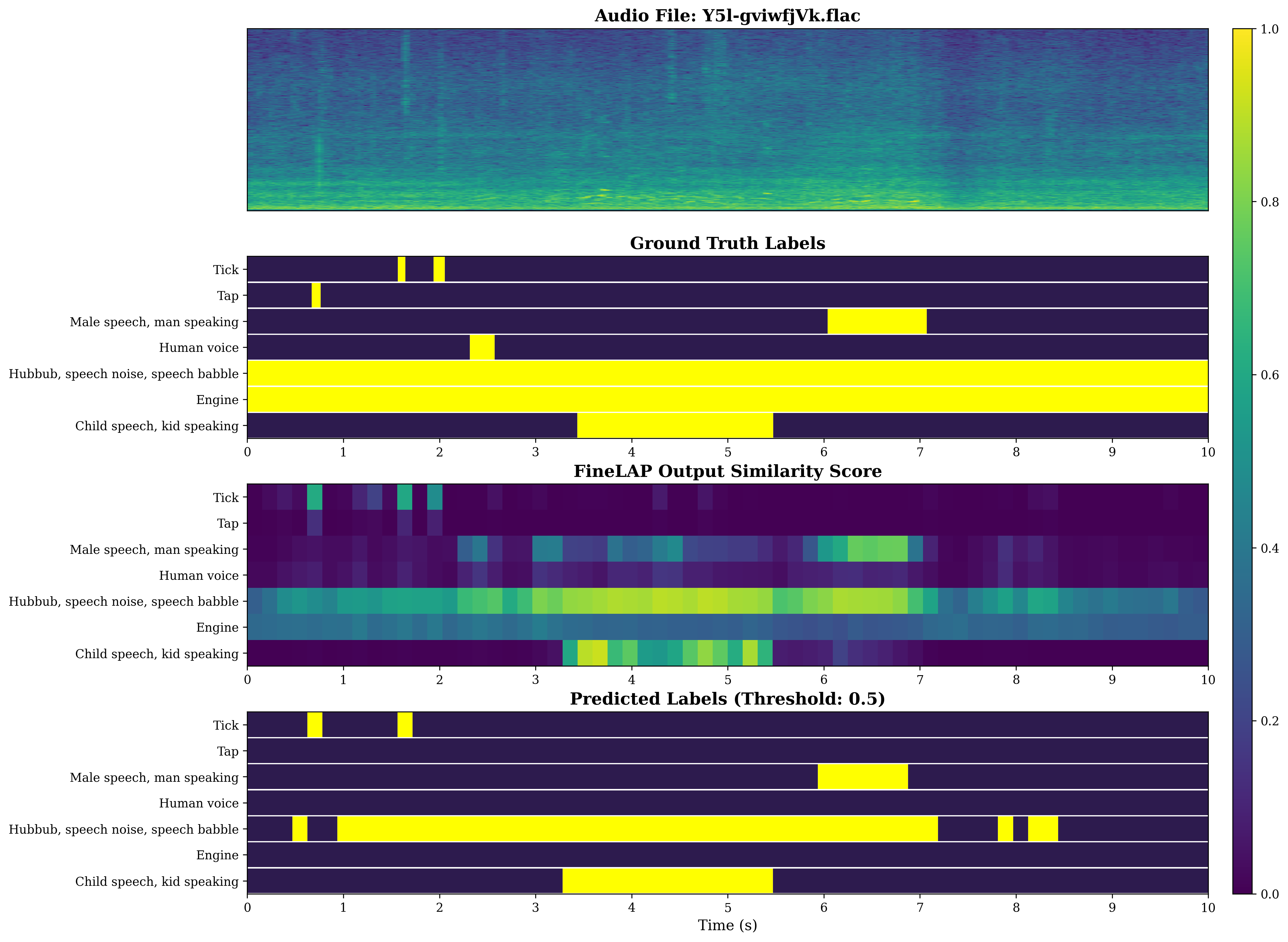}
    \vspace{0.5cm}
    \includegraphics[width=0.49\linewidth]{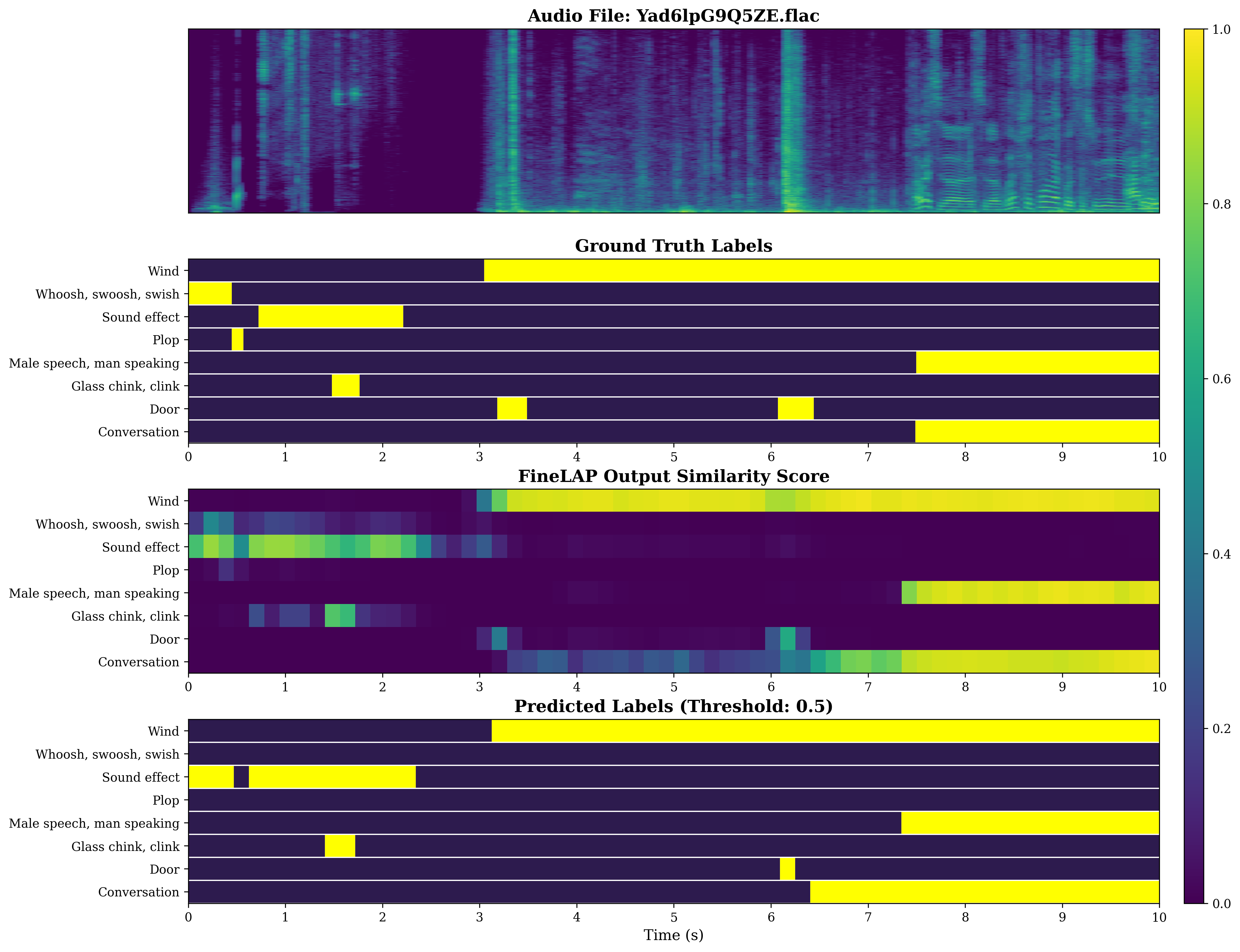} \hfill
     \includegraphics[width=0.49\linewidth]{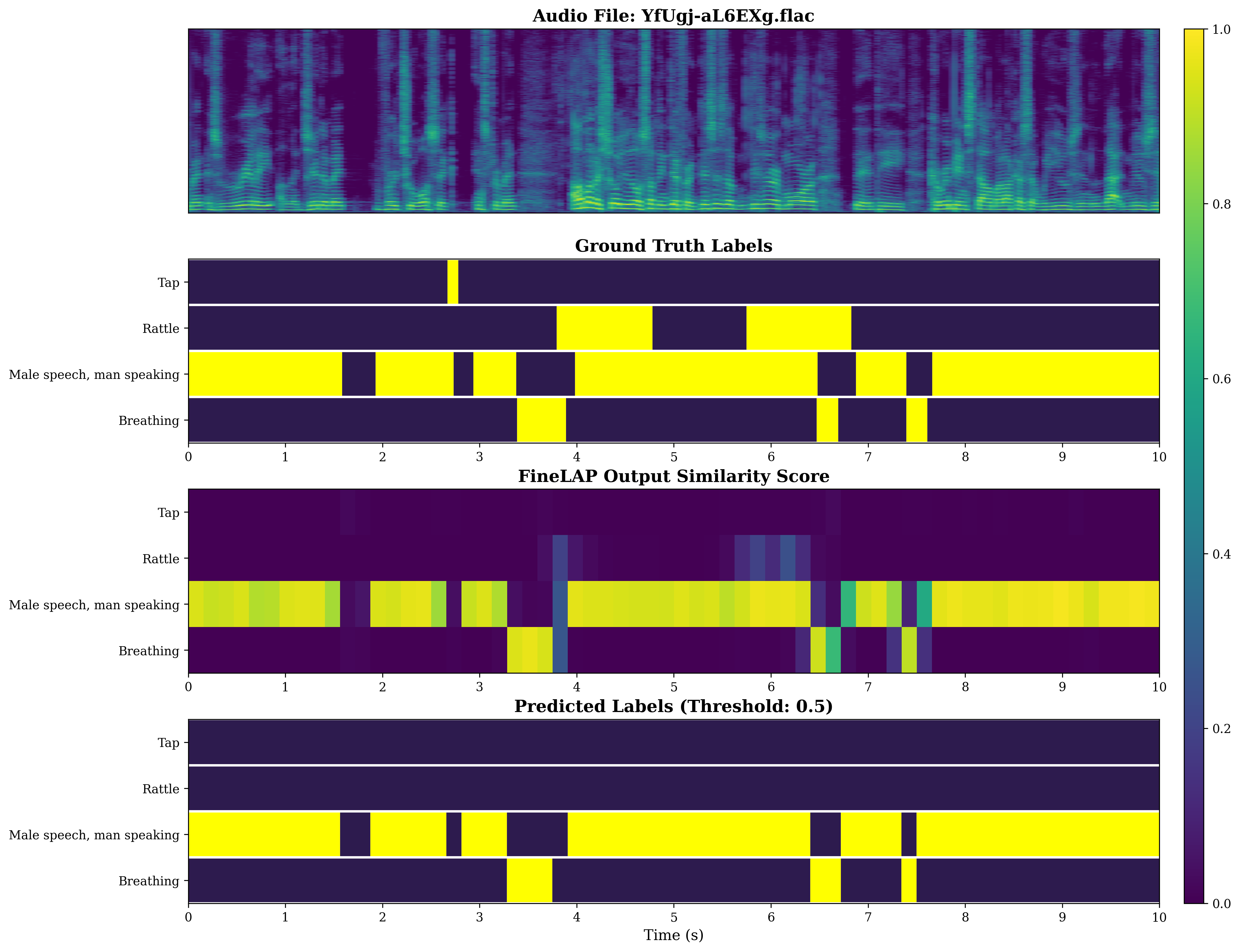}
    \caption{Visualization of FineLAP's performance on AudioSet-Strong }
\end{figure*}

\subsection{Window-based Audio Clipping Strategy}
\label{app:visualization_window-based_clipping}
\begin{figure*}
    \centering
    \includegraphics[width=0.49\linewidth]{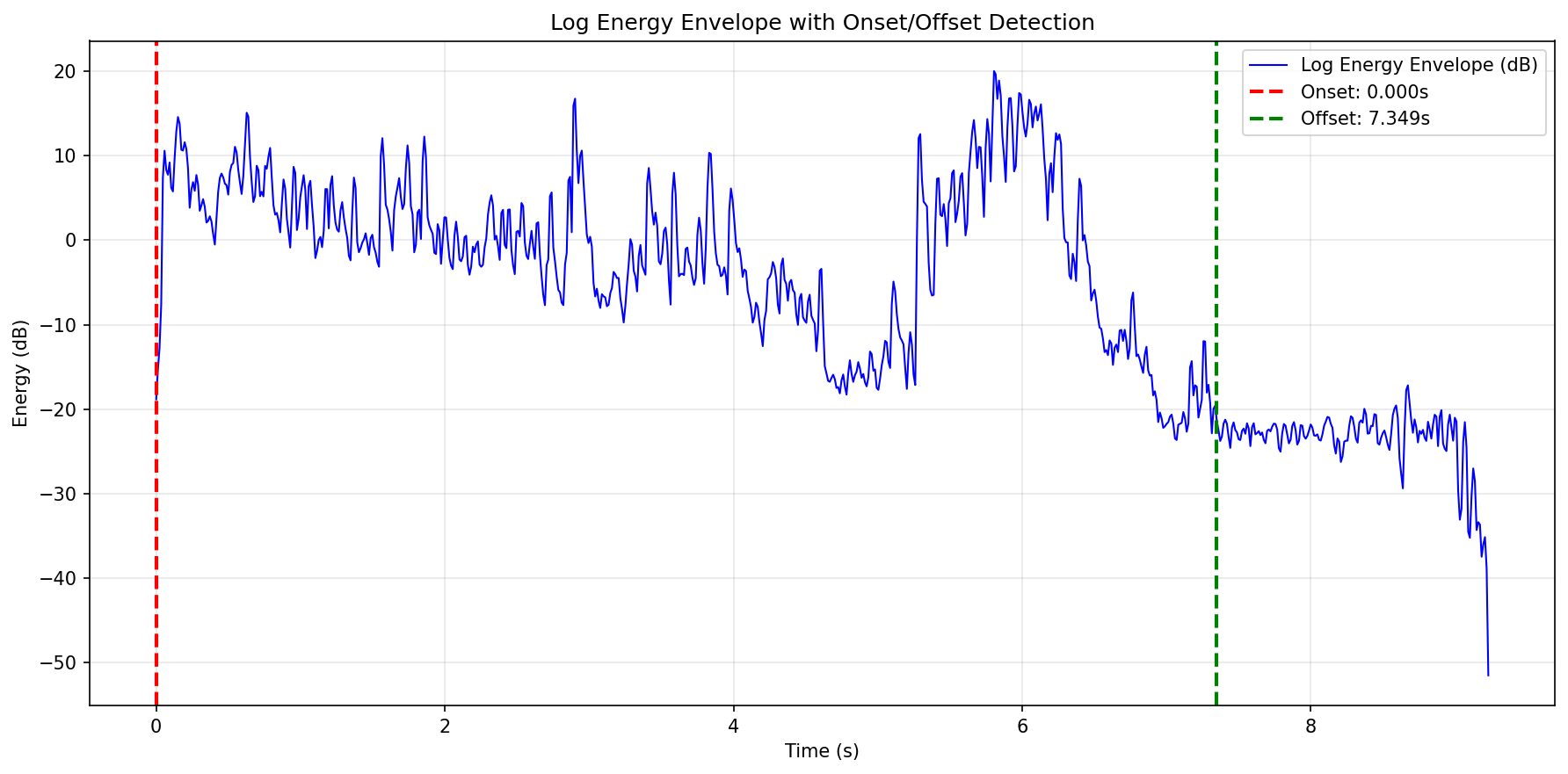} \hfill
    \includegraphics[width=0.49\linewidth]{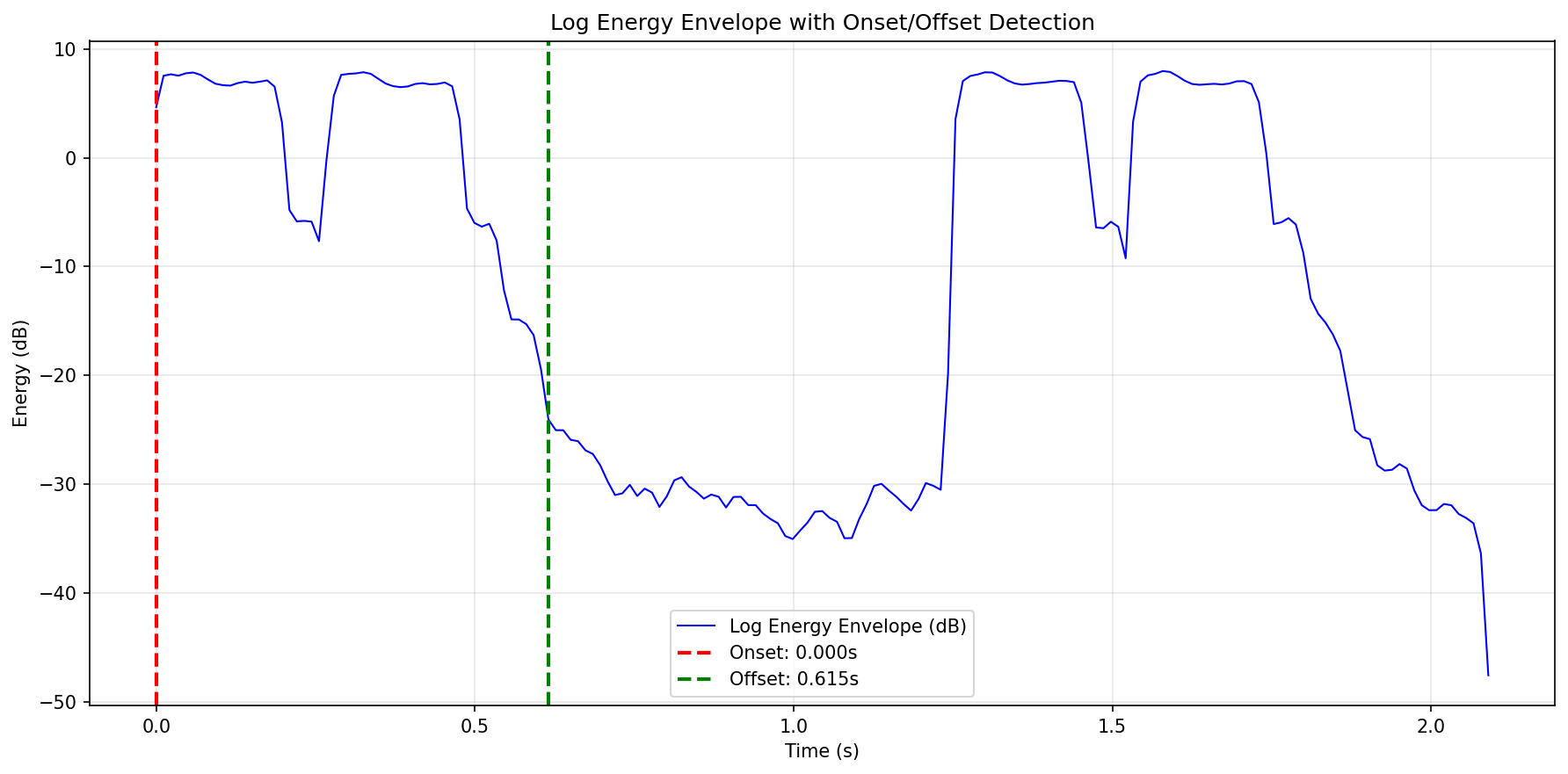}
      \vspace{0.5em}
    \includegraphics[width=0.49\linewidth]{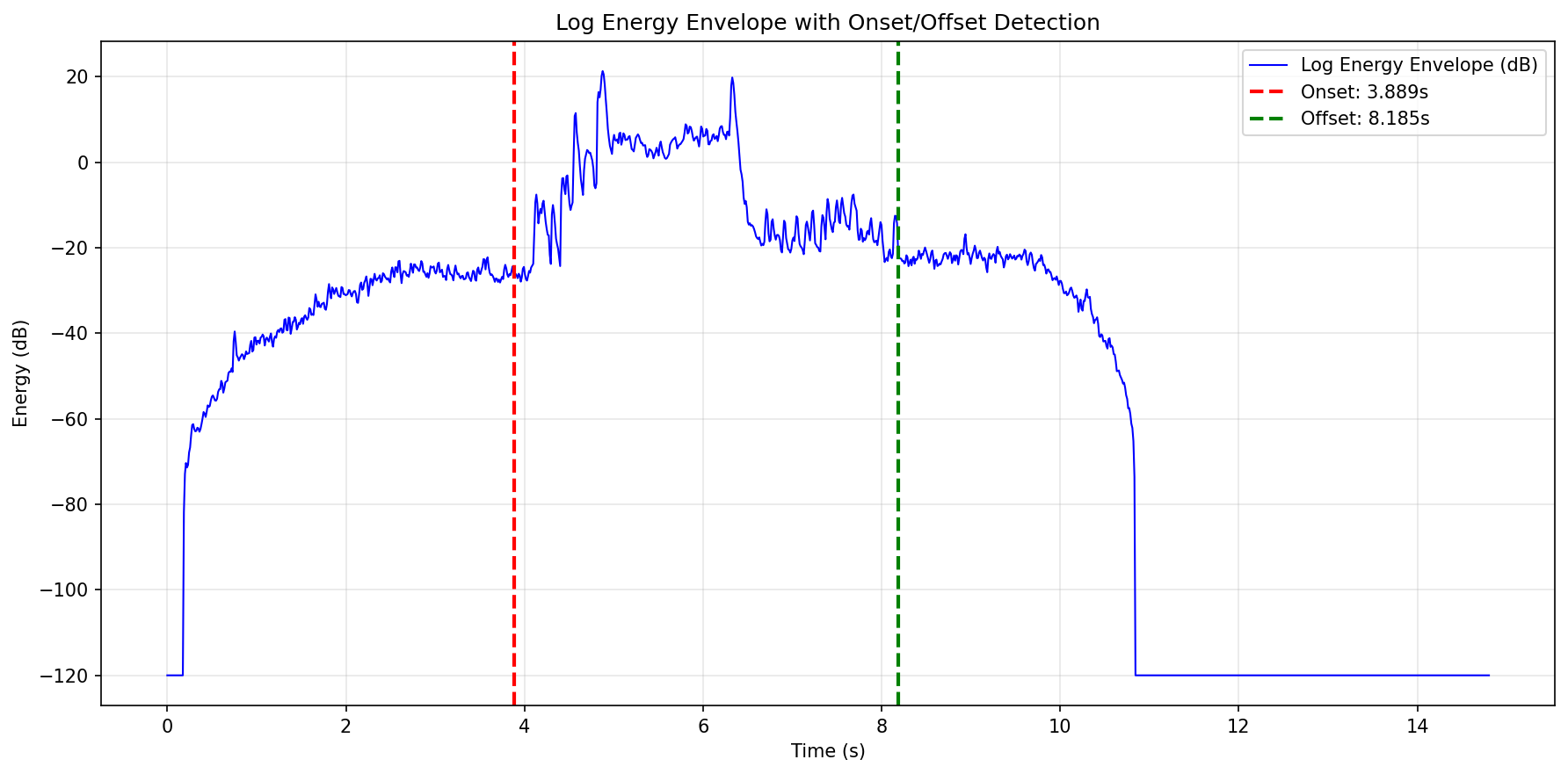}
     \includegraphics[width=0.49\linewidth]{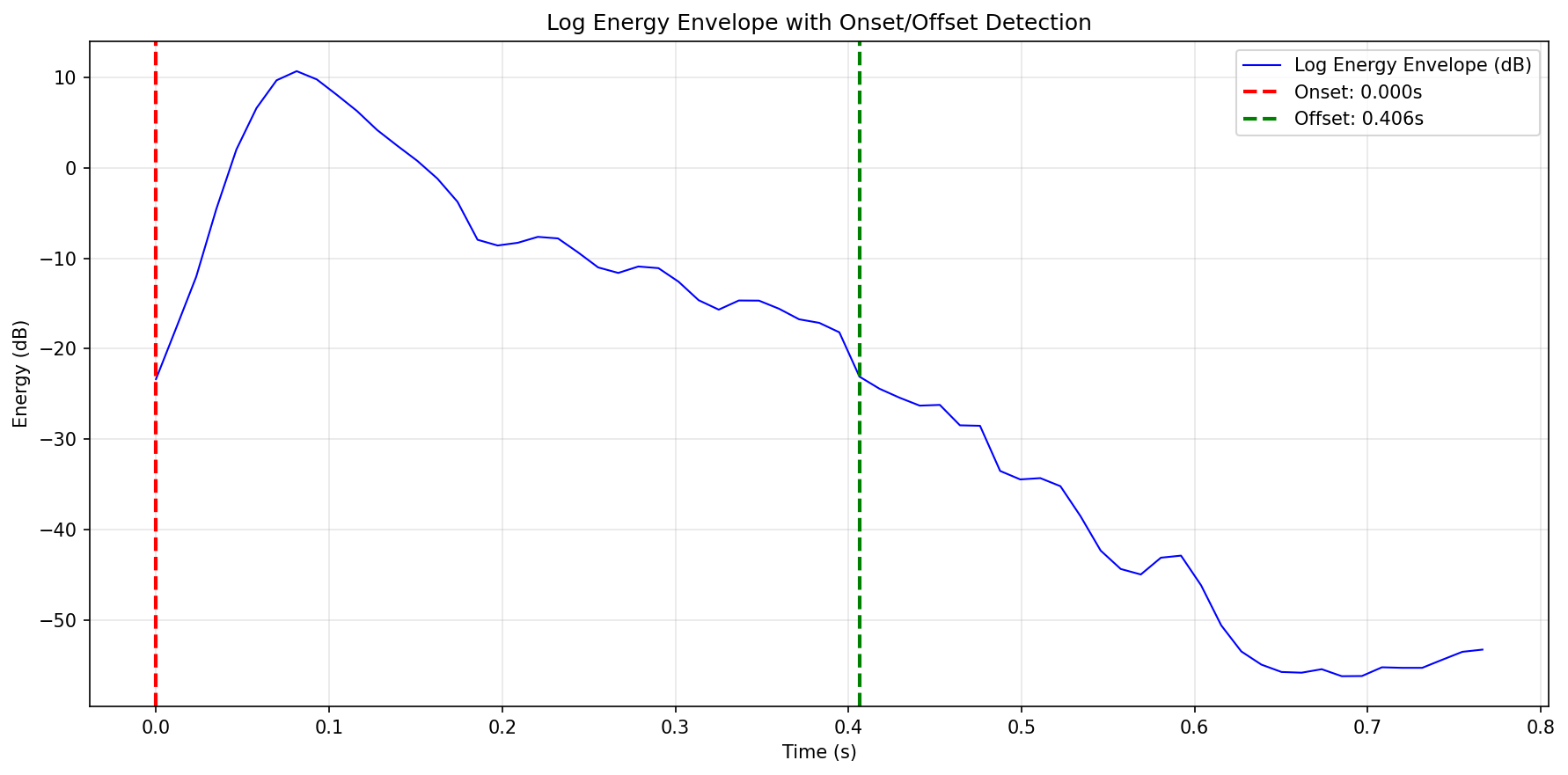}
    \caption{Visualization of the proposed window-based audio clipping strategy }
    \label{fig:window_based_audio_clipping}
\end{figure*}

As discussed in Section \ref{sec:data_curation_pipleine}, we introduce an energy window based audio clipping strategy to extract pure audio clip from real-world audio datasets. 
Here we present visualizations of our proposed strategies. 
As illustrated in Figure~\ref{fig:window_based_audio_clipping}, our strategy effectively identifies regions containing acoustic events and accurately crops clean audio segments by removing silent portions.

\section{Baseline Descriptions}
\label{app:baseline_descriptions}
\paragraph{LAION-CLAP. }
LAION-CLAP \cite{wu2023large} is one of the first large-scale pretrained audio-language models. The model adopts HTSAT and RoBERTa as its audio and text encoder. The authors employ keyword-to-caption data augmentation strategy along with feature fusion to improve model's capability. 

\paragraph{HTSAT-BERT. }
HTSAT-BERT \cite{mei2024wavcaps} is a strong audio-language model trained on the WavCaps dataset. It adopts HTSAT and BERT as its audio and text encoders. 

\paragraph{Cacophony. }
Cacophony \cite{zhu2024cacophony} is an improved contrastive audio-text model. It uses a self-supervised audio encoder trained with an MAE objective, and adopts a contrastive-captioning objective to enhance model's performance. 
Trained with a large-scale audio-caption dataset, Cacophony achieves strong performance across a wide range of audio understanding tasks. 

\paragraph{M2D-CLAP. }
M2D-CLAP \cite{niizumi2025m2d} is a strong audio-language model which extends an SSL masked modeling duo (M2D) \cite{niizumi2023masked} by incorporating CLAP and utilizes LLM-based sentence embeddings. It employs a multi-stage training pipeline to benefit from both self-supervised learning and contrastive language-audio pretraining. 

\paragraph{MGA-CLAP.}
MGA-CLAP \cite{li2024advancing} represents one of the pioneering works to advance fine-grained audio-text alignment in CLAP. It uses a modality-shared codebook, a temporal enhanced HTSAT encoder, along with a negative-guided contrastive loss to enhance both coarse- and fine-grained audio-language alignment. 
Despite achieving notable improvements over CLAP on tasks such as SED and grounding, MGA-CLAP does not explicitly leverage frame-level supervision, which results in suboptimal performance compared to specialized SED models.

\paragraph{FLAM. }
FLAM~\cite{wu2025flam} is the work most closely related to ours. FLAM introduces a frame-level contrastive objective to extend CLAP for open-vocabulary sound event detection (SED), and further incorporates a bias correction term together with an unbiased event classifier to address label imbalance during SED training. It dramatically improves CLAP's performance on open-vocabulary SED. 

In comparison, our work differs from FLAM in several key aspects:
\textbf{1). Training paradigm:}
FLAM adopts a multi-stage training strategy, where the model is first trained with a global contrastive objective on audio–caption data without frame-level annotations, and is then further optimized using a frame-level contrastive loss to enhance open-vocabulary SED performance.
In contrast, we aim to simultaneously advance both clip-level and frame-level audio–text alignment by jointly training on heterogeneous data sources. Empirically, we find that learning these two levels of alignment together leads to mutual benefits across tasks.
\textbf{2). Objective design:}
We replace the vanilla InfoNCE-based global contrastive loss with a sigmoid loss, and further introduce cluster-based negative sampling in the frame-level objective to improve semantic discrimination and representation learning.
\textbf{3). Model architecture:}
We propose a decoupled audio adapter built on top of a self-supervised audio encoder, which is shown to be beneficial for both clip-level and frame-level tasks by aligning multi-granularity representations into a shared embedding space.

\paragraph{PE$_\text{A-Frame}$. } 
PE$_\text{A-Frame}$ \cite{vyas2025pushing} is a concurrent work that focuses on frame-level audio–language alignment. It extends the Perception Encoder (PE) \cite{bolya2025perception} to a contrastive pre-training paradigm and scales training to $O$(100M) audio–video–text pairs, resulting in PE$\text{AV}$. Building on PE$\text{AV}$, PE$_\text{A-Frame}$ is further fine-tuned with a frame-level objective to enable temporally aligned audio–language understanding.

\paragraph{FlexSED}
FlexSED \cite{hai2025flexsed} is the state-of-the-art open-vocabulary sound event detection system. The model incorporated Dasheng \cite{dinkel2024scaling} as its audio encoder and introduces AdaLN-One to inject textual information from CLAP to obtain frame-level predictions. The model is trained on the AudioSet-Strong dataset. 

\paragraph{PretrainedSED}
PretrainedSED \cite{schmid2024effective} is the state-of-the-art closed-vocabulary sound event detection system. It employs a balanced sampler, aggressive data augmentation, and ensemble knowledge distillation to enhance model's performance. 
We report its performance of the distilled best model, which is based on the BEATs encoder \cite{chen2022beats}. 

\section{Training Data Details}
\label{app:training_data_details}

In this section, we present a detailed description of the datasets used in the training of FineLAP.

\noindent\textbf{AudioCaps. }AudioCaps \cite{kim2019audiocaps} contains approximately 50k audio-text pairs, where each audio contains one human-labeled high-quality caption.

\noindent\textbf{Clotho. } Clotho \cite{drossos2020clotho} contains $\sim$5k audio clips. Each audio is paired with 5 different human labeled captions. 

\noindent\textbf{WavCaps. }
WavCaps \cite{mei2024wavcaps} comprises 400k audio samples collected from multiple sources, including BBC Sound Effects,
\footnote{\url{https://sound-effects.bbcrewind.co.uk}}, 
FreeSound
\footnote{\url{https://freesound.org}}
, 
SoundBible
\footnote{\url{https://soundbible.com}} 
and AudioSet-Strong \cite{hershey2021benefit}.
Each audio has one caption generated by ChatGPT. 

\noindent \textbf{AudioSet. }
AudioSet \cite{gemmeke2017audio} contains approximately 2M audio samples. However, these audio segments only contain ground-truth labels. We use the caption from AudioSetCaps \cite{bai2025audiosetcaps} to train our CLAP model. 

\noindent \textbf{AudioSet-Strong. }
AudioSet-Strong \cite{hershey2021benefit} is a subset of AudioSet which contains temporally strong annotations. It is also the largest audio dataset which contains frame-level annotations. In total, it has 100k audio clips. 

\noindent \textbf{DESED-Strong. } 
DESED-Strong \cite{serizel2020sound} is another human-labeled sonud event detection dataset containing approximately 3k human labeled data. The dataset focuses on domestic sound events, with a total of 10 distinct labels. 

\noindent \textbf{UrbanSED. }
UrbanSED is a synthetic sound event detection dataset. It contains a total of 8k audio files, where each audio is synthesized using the Scaper \cite{salamon2017scaper} library. The original sound event files are taken from UrbanSound-8K. 


\section{Evaluation Details}
\label{app:evaluation_details}
Following prior work~\cite{hai2025flexsed, schmid2024effective, li2024self, vyas2025pushing}, we adopt the standard PSDS evaluation protocol \footnote{\url{https://github.com/fgnt/sed_scores_eval}}. 
For PSDS1, we set $\mathrm{DTC}=0.7$, $\mathrm{GTC}=0.7$, $\alpha_{\mathrm{ST}} = 1$, $\alpha_{\mathrm{CT}} = 0$, and $e_{\max} = 100$.
Consistent with~\cite{vyas2025pushing, hai2025flexsed}, we omit the variance penalty by setting $\alpha_{\mathrm{ST}} = 0$ for AudioSet-Strong, as PSDS was originally designed for datasets with fewer and less imbalanced classes.

\section{Algorithm}
We present a pseudo-code for our cluster-based negative sampling strategy in Algorithm~\ref{alg:cluster_negative_sampling_fixedN}. 

\begin{algorithm*}[t]
\caption{\textbf{Cluster-Based Negative Sampling with Fixed Phrase Set Size}}
\label{alg:cluster_negative_sampling_fixedN}
\begin{algorithmic}[1]
\State \textbf{Input:} Phrase database $\mathcal{D}_P$, phrase-to-cluster mapping $\mathcal{C}(\cdot)$, batch $\{(A^{(i)}, T^{(i)}, \{(P^{(i)}_k, Y^{(i)}_k)\}_{k=1}^{K_i})\}_{i=1}^{B}$, total phrase count $N$
\State \textbf{Output:} Enriched annotation sets $\{(P^{(i)}_k, Y^{(i)}_k)\}_{k=1}^{N}$ for frame-level supervision

\For{each audio clip $A^{(i)}$ in the batch}
    \State \textbf{Positive phrases:} $\mathcal{P}_i^{+} \gets \{P^{(i)}_k\}_{k=1}^{K_i}$
    \State \textbf{Positive clusters:} $\mathcal{C}_i^{+} \gets \{\mathcal{C}(P)\mid P \in \mathcal{P}_i^{+}\}$
    \State \textbf{Negative clusters:} $\mathcal{C}_i^{-} \gets \mathcal{C} \setminus \mathcal{C}_i^{+}$
    \State \textbf{Candidate negative pool:} $\mathcal{N}_i \gets \{P \in \mathcal{D}_P \mid \mathcal{C}(P) \in \mathcal{C}_i^{-}\}$
    \State Sample $N-K_i$ negative phrases: $\{P^{(i)-}_m\}_{m=1}^{N-K_i} \sim \mathcal{N}_i$
    \State \textbf{Enriched phrases:} $\{P^{(i)}_k\}_{k=1}^{N} \gets \mathcal{P}_i^{+} \cup \{P^{(i)-}_m\}_{m=1}^{N-K_i}$
    \State \textbf{Assign labels:} for each sampled negative phrase $P^{(i)-}_m$, set $Y^{(i)}[l]=0$ for all frames $l$
\EndFor

\end{algorithmic}
\end{algorithm*}

\end{document}